\documentclass[aps,prc,superscriptaddress,twoside,twocolumn,nofootinbib,showpacs,floatfix]{revtex4-1}
\usepackage{amsmath,amssymb}
\usepackage{graphicx}
\allowdisplaybreaks
\newcommand*{\fs}[1]{{#1\!\!\!/}}
\newcommand*{\hc}{\text{H.\,c.}}

\begin{document}

\title{\boldmath Combined analysis of $\eta'$ production reactions: $\gamma N \to \eta' N$, $NN \to NN\eta'$, and $\pi N \to \eta'N$}

\author{F. Huang}
\email{huang@physast.uga.edu}
\affiliation{Department of Physics and Astronomy, The University of Georgia, Athens, GA 30602, USA}

\author{H. Haberzettl}
\email{helmut.haberzettl@gwu.edu}
\affiliation{Institute for Nuclear Studies and Department of Physics, The George Washington University, Washington, DC 20052, USA}

\author{K. Nakayama}
\email{nakayama@uga.edu}
\affiliation{Department of Physics and Astronomy, The University of Georgia, Athens, GA 30602, USA}
\affiliation{Institut f{\"u}r Kernphysik (Theorie) and J\"ulich Center for Hadron Physics, Forschungszentrum J{\"u}lich, 52425 J{\"u}lich, Germany}

\date{\today --- \jobname}

\begin{abstract}
The production of $\eta'$ mesons in photon- and hadron-induced reactions has
been revisited in view of the recent additions of high-precision data to the
world data base. Based on an effective Lagrangian approach, we have performed a
combined analysis of the free and quasi-free $\gamma N \to \eta' N$, $NN \to
NN\eta'$, and $\pi N \to \eta' N$ reactions. Considering spin-1/2 and -3/2
resonances, we found that a set of above-threshold resonances $\{S_{11}$,
$P_{11}$, $P_{13}\}$, with fitted mass values of about $M_R=1925$, $2130$, and
$2050$ MeV, respectively, and the four-star sub-threshold $P_{13}(1720)$
resonance reproduce best all existing data for the $\eta'$ production processes
in the resonance-energy region considered in this work. All three above-threshold resonances found in
the present analysis are essential and indispensable for the good quality of
the present fits.
\end{abstract}


\maketitle


\section{Introduction}\label{sec:Intro}

A wealth of interesting physics can be obtained from studying the production
processes involving the $\eta^\prime$ meson \cite{NH04}, one of the primary
interests being here that such production processes may help one extract
information on nucleon resonances
that cannot be obtained from pion reactions. In fact, current knowledge of most
of the nucleon resonances is mainly due to the study of $\pi N$ scattering
and/or pion photoproduction off the nucleon. Since the $\eta'$ meson is much
heavier than the pion, $\eta'$ meson-production processes near threshold,
therefore, are well suited for investigating high-mass resonances in low
partial-wave states. Furthermore, reaction processes such as $\eta'$
photoproduction provide opportunities to study, in particular, those resonances
that couple only weakly to pions. This may help in providing a better
understanding of the so-called ``missing resonances'' predicted by quark
models \cite{Capstick1}, but not found in more traditional pion-production reactions.

In view of the relatively low production rate, until recently there
existed only a very limited number of experimental studies of
$\eta^\prime$ production reactions. This limited experimental information was
reflected in the relatively low number of theoretical investigations of such
reactions. For an account of the pre-2004 situation, see Ref.~\cite{NH04}.
However, the situation has changed in the past few years, especially in
$\eta^\prime$ photoproduction, where high-precision data for both nucleon and
deuteron targets have become available \cite{CLAS06,CLAS09, CBELSA09,CBELSA11}.
Also, the $pp$ and $p\eta^\prime$ invariant-mass distribution data in the $pp
\to pp\eta^\prime$ reaction are now available \cite{COSY11-10a}, in addition to
the cross-section data \cite{ppetapdata,Balestra,Khoukaz}. Upper limits for the
total cross sections in $pn\to pn\eta'$ have also been reported
\cite{COSY11-10b}. Given the present situation, with much higher-quality data
than were available in the past, we revisit here the production of
$\eta^\prime$ and we perform a combined analysis of the reaction
channels  $\gamma N \to \eta' N$, $NN \to NN\eta'$, and $\pi N \to\eta'N$.
In our previous study of $\eta^\prime$ photoproduction on the proton
\cite{NH04,NH06}, we found that cross-section data alone cannot unambiguously
constrain the set of minimal spin-1/2 and -3/2 resonances necessary in
principle for an adequate reproduction of the data, and we pointed to the
necessity of incorporating spin observables (in particular, the beam asymmetry)
to constrain our model much more stringently than what can be achieved by
utilizing the cross section alone.  To date, although there are ongoing efforts to 
measure the beam asymmetry \cite{Collins09} and the beam-target asymmetry 
\cite{Afzal12}, experimental data
for spin observables are not available yet for $\eta^\prime$ production
reactions. However, as we shall show in this work, the \emph{simultaneous}
consideration of the high-precision cross-section data in free $\gamma p \to
\eta^\prime p$ \cite{CLAS09} that have only become available recently, together
with the cross section data on quasi-free $\gamma n \to \eta^\prime n$
\cite{CBELSA11} and the invariant mass distribution data in $pp \to
pp\eta^\prime$ \cite{COSY11-10a}, impose sufficient constraints to remove the
ambiguity among various sets of possible spin-1/2 and -3/2 resonances. 

To analyze the new photoproduction data \cite{CLAS09,CBELSA09,CBELSA11}, we use
here the same approach as employed in Ref.~\cite{NH06} for the analysis of the
earlier CLAS data \cite{CLAS06}. In particular, we include spin-1/2 and -3/2 resonances
with parameters determined from best fits to the data. We restrict ourselves to these low-spin
resonances because the existing cross-section data alone cannot constrain the resonance 
parameters unambiguously once higher-spin resonances are added into the model. For this,
one must have spin-observable data. Since the
present approach is phenomenological, our strategy is to consider the minimum
number of resonances that allows us to reproduce the available data with
acceptable fit accuracy. This is quite different in spirit from a recent
analysis of the new data \cite{CLAS09,CBELSA09,CBELSA11} by Zhong and Zhao
\cite{ZZ11} who work within a quark-model approach and consider all possible
nucleon resonances up to the $n=3$ harmonic-oscillator shell. For the
quasi-free photoproduction processes \cite{CBELSA11}, we account for the Fermi
motion of the nucleon by folding the cross section of the free process with the
momentum distribution of the nucleon inside the deuteron. The analysis of the
reaction $NN \to NN\eta'$ is done following Ref.~\cite{NH04} within a
distorted-wave Born approximation (DWBA) in which both the initial- and
final-state $NN$ interaction is taken into account explicitly.

This paper is organized as follows: In Sec.~\ref{sec:results}, we describe our overall
strategy for performing our combined analysis of the photon- and hadron-induced
reactions and we also provide
some general remarks concerning the resonances required to reproduce the
available data. In Sec.~\ref{sec:free-proton1}, the present results for the
free $\gamma p \to \eta' p$ reaction are discussed in conjunction with the most
recent high-precision CLAS \cite{CLAS09} and CBELSA/TAPS \cite{CBELSA09} data.
In Sec.~\ref{sec:quasi-free}, the analysis of the quasi-free $\gamma p \to
\eta^\prime p$ and $\gamma n \to \eta^\prime n$ reactions is presented. Section
\ref{sec:NNeta'} contains the present results for the $NN \to NN\eta^\prime$
reaction and in Sec.~\ref{sec:eta'N} we present the results for the
pion-induced $\eta'$ production $\pi N\to \eta' N$. Our summarizing assessment
is given in Sec.~\ref{sec:conclusions}. In Appendix \ref{sec:appA}, our model
for photoproduction as well as for the hadron-induced reactions $NN \to
NN\eta'$ and $\pi N \to \eta' N$ are described briefly for completeness. Some
details of how the Fermi motion of the nucleon inside the deuteron is taken
into account in the present work for describing the quasi-free photoproduction
processes are also given in Appendix \ref{sec:appA}. Appendix \ref{sec:appB}
contains the Lagrangians, form factors, and propagators that define the
individual amplitudes in our model.

\section{General Procedure and Findings}\label{sec:results}

In the following sections, we present separate discussions of our results for
photon- and  for hadron-induced reactions because this allows us to keep the
discussions focused on the reactions at hand. We emphasize, however, that the
results are based on \emph{simultaneous} fits to the available data for the
reaction processes considered.

The model assumptions used in describing these reactions consistently with each
other are given in Appendix \ref{sec:appA}. In addition to resonance-current
contributions, we consider the nucleonic and meson-exchange ($\rho$ and
$\omega$) currents. For resonances, in particular, the strategy adopted here is
to introduce as few resonances as possible, with parameters adjusted to
simultaneously reproduce the data for the $\gamma N \to \eta' N$, $NN \to
NN\eta'$, and $\pi N \to \eta' N$ reactions via a least-square minimization
procedure. In the present work, we restrict ourselves to spin-1/2 and -3/2
resonances.

The simultaneous treatment of the available reaction data provides insights
that cannot be obtained by fitting each reaction and/or each data set
individually. For example, considering only the free $\gamma p \to \eta' p$
photoproduction data from the CBELSA/TAPS experiments \cite{CBELSA09}, we can
obtain good fits (with $\chi^2/N \sim 1$) with only \emph{one} resonance. By
contrast, a similarly good fit to the corresponding high-precision CLAS data
\cite{CLAS09} requires at least \emph{three} resonances. Considering two
above-threshold resonances instead, one obtains $\chi^2/N=1.7$ at best.
Moreover, we find that any one of the above-threshold three-resonance sets
$\{P_{11}, P_{13}, D_{13}\}$, $\{S_{11}, P_{11}, P_{13}\}$, and $\{S_{11},
P_{13}, D_{13}\}$ can reproduce the data equally well. The remaining
combination $\{S_{11}, P_{11}, D_{13}\}$ is ruled out by the data according to
our good-fit criterion thus indicating the need for the $P_{13}$ resonance for
reproducing the data. These findings indicate that in $\gamma p \to \eta' p$
the cross-section data alone cannot constrain the set of resonances, even with
high-precision data. By contrast, the three acceptable sets of resonances given
above yield quite distinct results for the spin observables, in particular, for
the beam asymmetry. These findings corroborate the conclusions of our earlier
work \cite{NH06}.

The ambiguity just discussed with respect to the three resonance sets found
acceptable in the free photoproduction process $\gamma p \to \eta' p$ is
completely removed once we include other reactions in our analysis. The set
$\{P_{11},P_{13},D_{13}\}$ is ruled out by the quasi-free $\gamma n \to \eta'
n$ reaction and the set $\{S_{11},P_{13},D_{13}\}$ by the $NN \to NN\eta'$
reaction. The only remaining acceptable set, therefore, is
$\{S_{11},P_{11},P_{13}\}$.
In addition,
we find that to obtain a good description of $NN \to NN\eta'$, in particular,
we need an additional \emph{below}-threshold resonance  whose inclusion has no
bearing on the quality of any of the results for other reactions considered in
this work. In summary, therefore, a good overall description of all reaction
processes considered in this work is obtained with a (minimum) set of three
above-threshold resonances, $\{S_{11},P_{11},P_{13}\}$, with the masses of
about $M_R=1925$, 2130, and 2050 MeV, respectively, and one below-threshold
$P_{13}$ resonance. The latter is the four-star $P_{13}(1720)$ quoted in PDG
\cite{PDG}. The masses of these above-threshold resonances are very well
constrained by the existing data, in particular, by the CLAS photoproduction
data \cite{CLAS09}. The CLAS data also constrain reasonably the total widths of
these resonances, although to a lesser extent. Hereafter, we shall use the
notation $S_{11}(1925), P_{11}(2130), P_{13}(2050)$ to identify these
above-threshold resonances, even though the fitted mass values for the various
scenarios discussed below (see Table \ref{tab:1}) may differ slightly from
these values. The resonances found in our analysis may be tentatively
identified with the corresponding $S_{11}(1895)$**, $P_{11}(2100)$*, and
$P_{13}(2040)$* resonances listed in PDG \cite{PDG}. We will discuss the
constraints imposed by the individual reactions in more detail in subsequent
sections.

\section{\boldmath Free $\gamma p \to \eta' p$} \label{sec:free-proton1}

In this section, we present our results for the free $\gamma p \to \eta^\prime
p$ reaction. First, we address the issue of possible discrepancies in the data
sets from the CLAS \cite{CLAS09} and the CBELSA/TAPS \cite{CBELSA09}
collaborations. Then, we discuss the present analysis of these data.

\begin{figure*}[tb]
\includegraphics[width=0.8\textwidth,angle=0,clip]{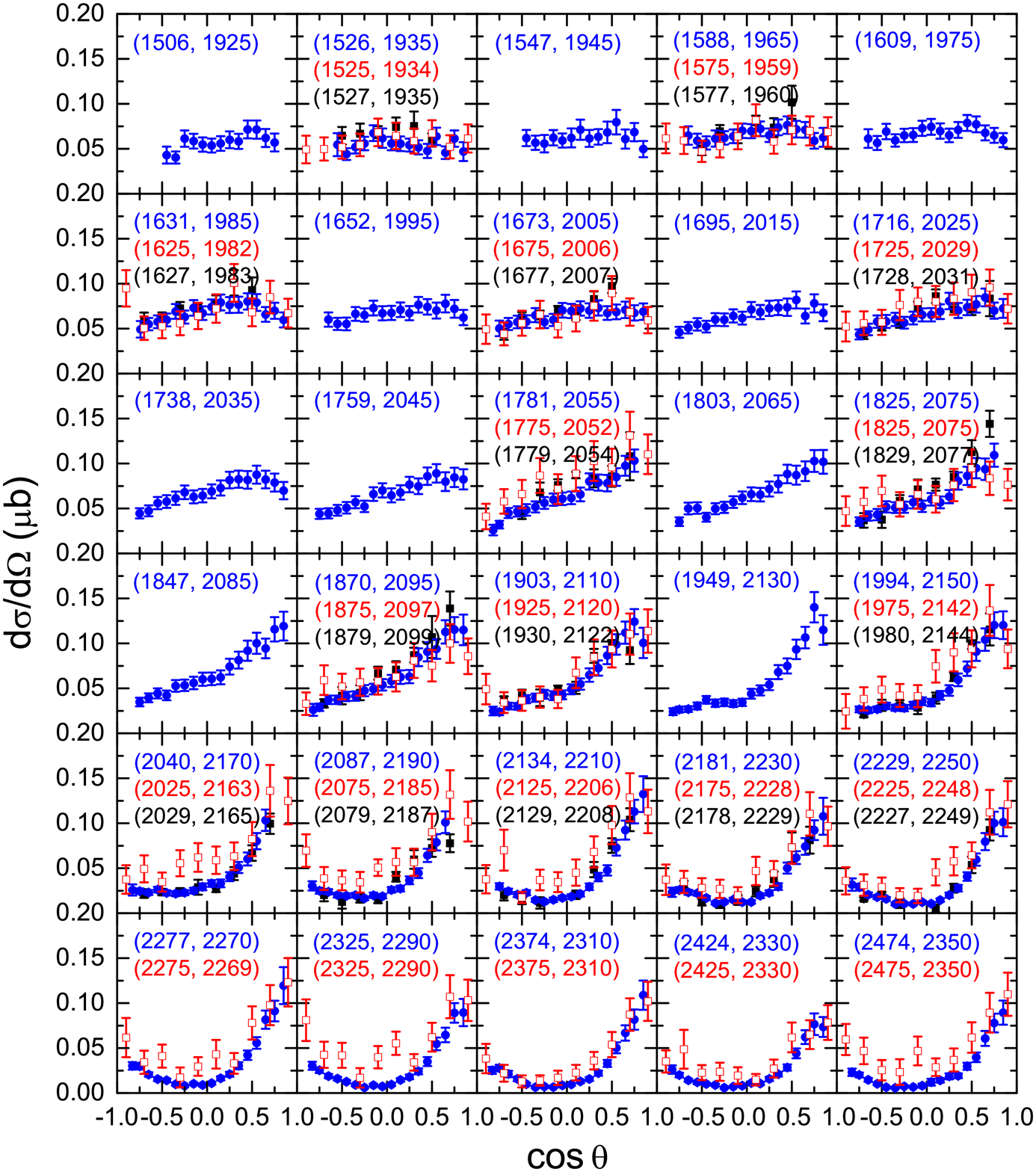}
\vskip -0.5cm \caption{\label{fig:1} (Color online) Comparison of the CLAS
\cite{CLAS09} (blue solid circle) and CBELSA/TAPS \cite{CBELSA09} (red open
square) differential cross-section data for the free $\gamma p \to \eta^\prime
p$ reaction as a function of $\cos\theta$ (where $\theta$ is the
$\eta^\prime$ emission angle in the center-of-momentum frame) for invariant energies up
to $\sqrt{s} = 2.35$ GeV. The earlier CLAS data \cite{CLAS06} (black solid
square) are also shown. The numbers in parentheses denote the photon laboratory
incident energy (left number) and the total energy (right number) of the
system, with the upper number pair pertaining to the newer CLAS data, the
second one to the CBELSA/TAPS experiment, and (where present) the lower one to
the older CLAS data. Note that the energies for different experiments shown in
the same panel are within $\pm 10$ MeV. }
\end{figure*}

\begin{figure*}[tb]
\includegraphics[width=0.8\textwidth,angle=0,clip]{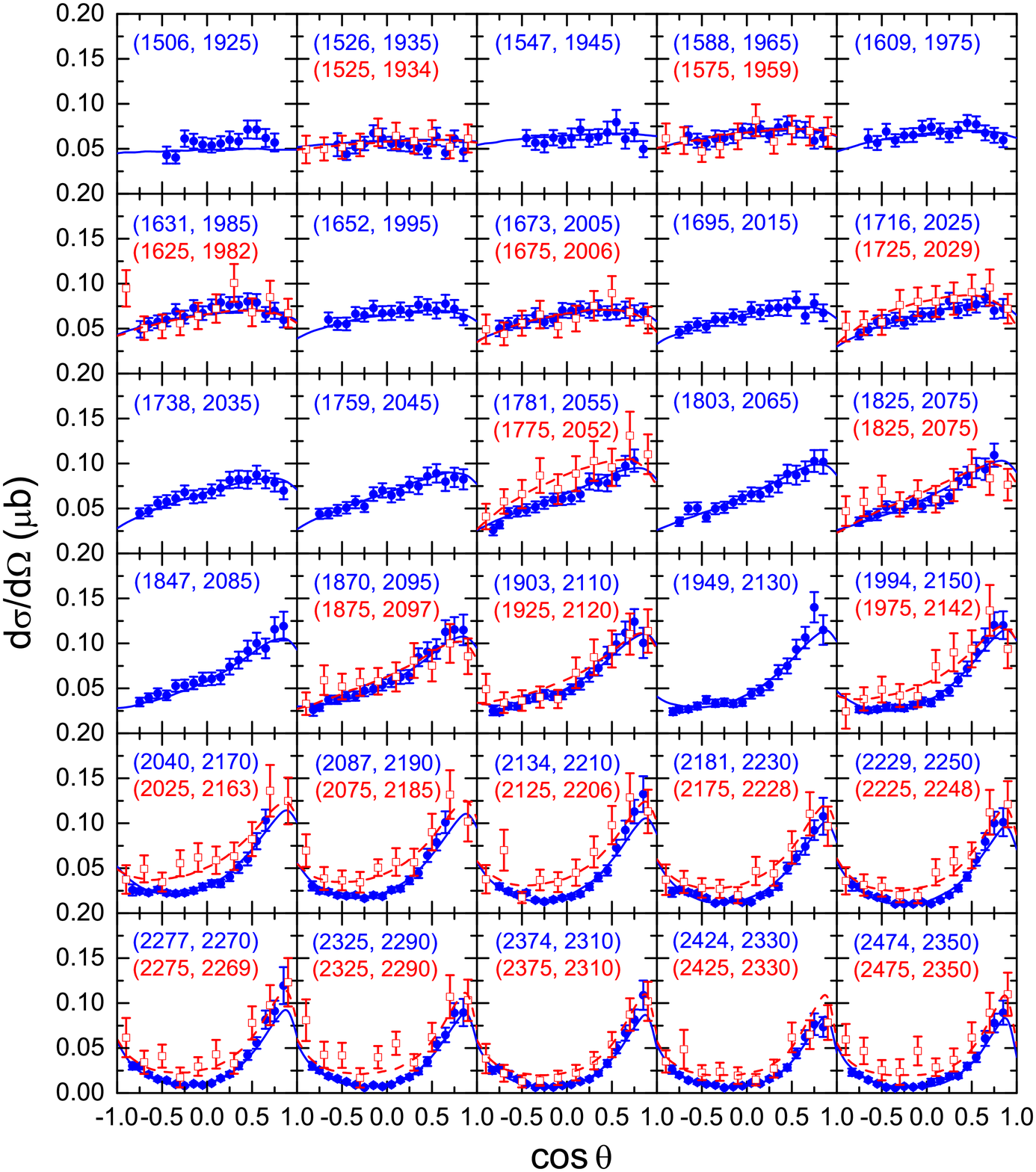}
\vskip -0.5cm
\caption{\label{fig:3} (Color online) Same differential cross-section data as Fig.~\ref{fig:1} (but without older CLAS data \cite{CLAS06}), now with curves resulting from our fits discussed in
Sec.~\ref{sec:free-proton2}. The CLAS data and fit curves are shown as blue
solid circles and blue solid curves, respectively, and the CBELSA/TAPS data and fit
curves are depicted as red open squares and red dashed curves, respectively.
The curves are the results of a combined fit of the photon- and hadron-induced
reactions data and both include the set of  resonances $\{P_{13}(1720), S_{11}(1925), P_{11}(2130), P_{13}(2050)\}$ with parameters given in Table \ref{tab:1}. All other parameter values that influence the
photoproduction reaction directly are also given in Table \ref{tab:1}. }
\end{figure*}

\begin{figure}[tb]
\includegraphics[width=1.0\columnwidth,angle=0,clip]{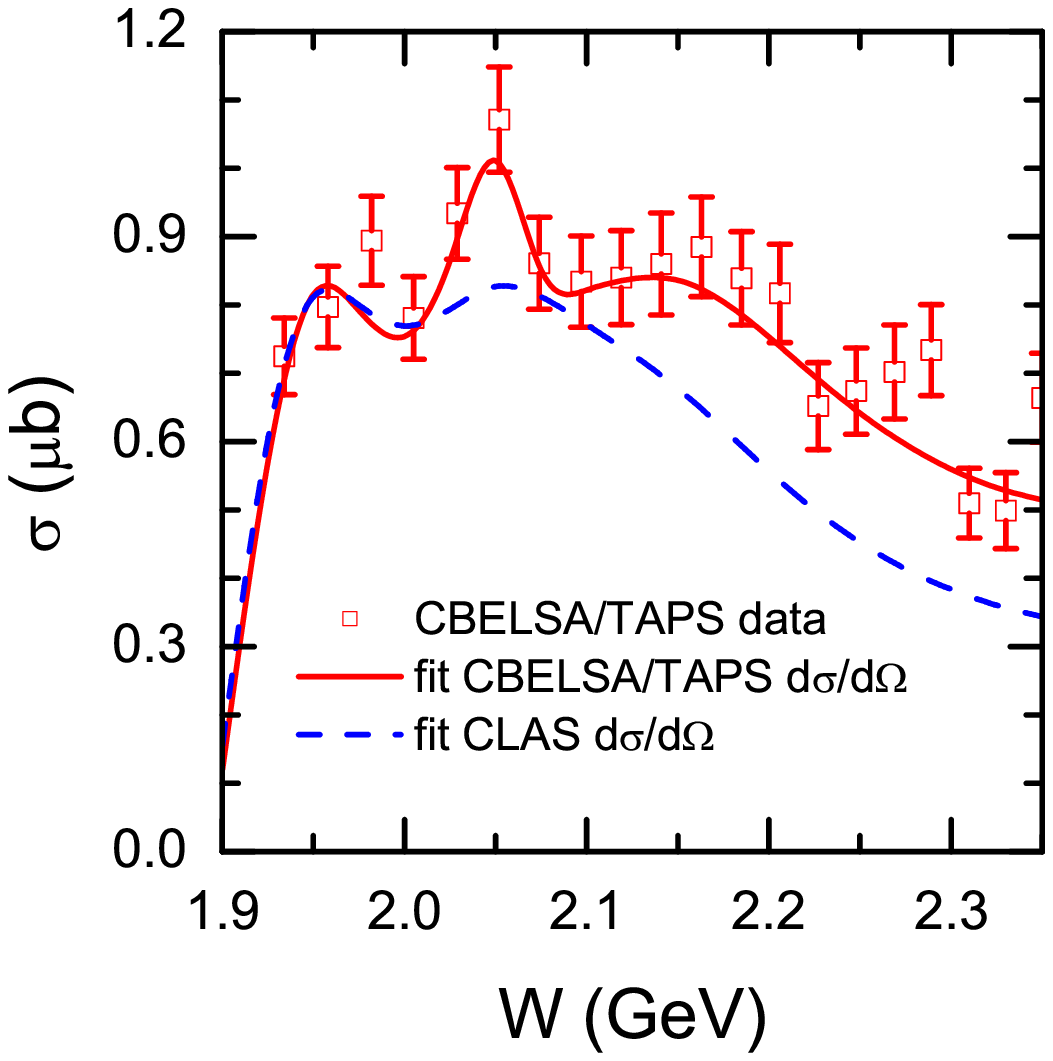}
\vskip -0.5cm
\caption{\label{fig:2}
(Color online) Predictions for the total cross sections in (free) $\gamma
p \to \eta' p$ obtained by integrating the corresponding fit results of the
CLAS \cite{CLAS09} (dashed blue line) and the CBELSA/TAPS \cite{CBELSA09}
(solid red line) angular distribution data as shown in
Fig.~\ref{fig:3} using the parameters of Table \ref{tab:1}. The data are from CBELSA/TAPS
\cite{CBELSA09} obtained by simply integrating the corresponding differential
cross section. They are not included in the fit. }
\end{figure}

\subsection{Comparison of the CLAS and CBELSA/TAPS data} \label{subsec:1}

The most recent CLAS data \cite{CLAS09} and the CBELSA/TAPS data
\cite{CBELSA09} are compared in Fig.~\ref{fig:1}. As one can see, in general
the new CLAS and the CBELSA/TAPS data are consistent for invariant energies
below $\sqrt{s} \sim 2.0$ GeV within their uncertainties. For higher energies,
however, one sees considerable discrepancies, by factors as large as 3, between
these two sets of data for $\eta^\prime$ emission angles away from the forward
angles. We note that in addition to the statistical errors, the plots of
Fig.~\ref{fig:1} include the estimated systematic errors as quoted in
Refs.~\cite{CLAS09,CBELSA09}.  These systematic errors were not included in the
data plots given in these references. The error bars in Fig.~\ref{fig:1} were
obtained by adding the systematic errors to the corresponding statistical
errors in quadrature. For the present purpose of revealing the inconsistencies
between these two data sets, this procedure may be justified, even though one
might need a more thorough error analysis for a fully quantitative estimate of
the total uncertainty in the data. We mention that for $\eta$ photoproduction,
an even more pronounced discrepancy between the CLAS and CBELSA/TAPS data was
pointed out in Refs.~\cite{CLAS09,CBELSA09}, however, with no clear
identification of the source of the discrepancy (in this respect, see also
Ref.~\cite{SHKM10}). For comparison, we show in Fig.~\ref{fig:1} also the
earlier CLAS data \cite{CLAS06} which are seen to be consistent with the newer
CLAS data.

Leaving out the older CLAS data \cite{CLAS06}, we show in Fig.~\ref{fig:3} the
same data as in Fig.~\ref{fig:1}, however, now with curves that result from our
fit procedure to the new CLAS \cite{CLAS09} and the CBELSA/TAPS \cite{CBELSA11}
data. We will discuss these fit results in detail in the subsequent
Sec.~\ref{sec:free-proton2}. Here, we note that if we integrate our fit results
for the corresponding angular distributions in Fig.~\ref{fig:3}, we obtain the
total cross sections shown in Fig.~\ref{fig:2} which exhibit markedly different
behavior for the two data sets. First, these results clearly reveal an
energy-dependent relative normalization problem which increases with increasing
energy. We checked, of course, how much this finding is influenced by the
uncertainties of the fit procedure and we found that other fits to the data of
similar quality, even polynomial fits, have no effect on our conclusion
regarding the normalization problem. Again, a similar finding was observed for
$\eta$ photoproduction data \cite{CLAS09,CBELSA09}, as was pointed out in
Ref.~\cite{SHKM10}. Second, the CBELSA/TAPS data exhibit a pronounced peak
structure around 2.05 GeV, while the total cross-section resulting from the fit
to the CLAS data shows only a relatively flat bump at this energy. The origin
of this pronounced peak structure can be traced back to the measured cross
sections in the CBELSA/TAPS data bin around 2.052 GeV. We will discuss this
structure in more detail in the following subsection.

\subsection{Analysis of the CLAS and CBELSA/TAPS data}\label{sec:free-proton2}

The discrepancy between the CLAS \cite{CLAS09} and CBELSA/TAPS \cite{CBELSA09}
data discussed in the previous subsection makes it difficult to consider both
data sets as a single set of data for a combined numerical
analysis. Since, at present, we have no clear reason to discard one data set in
favor of the other, we are forced, therefore, to consider them separately. In
view of their differences, we can expect that the resonance parameters
extracted from these data sets may be quite different from each other. One of
the purposes of this subsection is to see how different they are.

\begin{table*}[tb]
\caption{\label{tab:1}Model parameter values that directly affect the
photoproduction reaction obtained in a combined analysis of the photon- and
hadron-induced reactions. The values in the columns labeled CLAS and CBELSA/TAPS
subsumed under ``free $p$" correspond to the fit results for the CLAS
\cite{CLAS09} and CBELSA/TAPS \cite{CBELSA09} free proton data, respectively.
The last row corresponds to the fit results for the CBELSA/TAPS quasi-free
proton data \cite{CBELSA11} discussed in Sec.~\ref{sec:quasi-free}.  Values in
boldface were kept fixed during the fitting procedure. (The fixed mass
values for the quasi-free calculation are obtained as averages of the
corresponding CLAS and CBELSA/TAPS values; see text for further explanation.) For
the definition of the parameters, see Appendices \ref{sec:appA} and
\ref{sec:appB}. The resonance mass and total width, $M_R$ and $\Gamma_R$,
are both in units of MeV, while the reduced helicity amplitude, $\sqrt{\beta_{N\eta'}}A_j$, is in units of
$10^{-3}$GeV$^{-1/2}$. }
\begin{center}
\begin{tabular*}{\textwidth}{@{\extracolsep\fill}lrrrrrrrrr}
\hline\hline
                    &&& \multicolumn{4}{c}{free $p$} &&& quasi-free $p$ \\ \cline{4-7} \cline{10-10}
                    &&& CLAS &&& CBELSA/TAPS \\  \hline
$\chi^2/N$          &&& $0.65$ &&& $0.53$ &&& $0.77$ \\ \hline
$g_{NN\eta'}$       &&& $1.00\pm 0.06$ &&& $1.17\pm 0.31$ &&& $1.00\pm 0.24$ \\
$\lambda_{NN\eta'}$ &&& $0.53\pm 0.06$ &&& $0.44\pm 0.22$ &&& $0.64\pm 0.24$ \\
$\Lambda_v$ [MeV]       &&& $1183\pm 5$ &&& $1244\pm 35$ &&& $1221\pm 28$ \\
$\hat{h}$           &&& $3.89\pm 0.18$ &&& $5.37\pm 1.57$ &&& $4.27\pm 0.89$ \\ \hline
$P_{13}(1720)$      &&&        &&&        && \\
$M_R        $          &&& ${\bf 1720}$ &&& ${\bf 1720}$ &&& ${\bf 1720}$ \\
$\Gamma_R   $         &&&  ${\bf 200}$ &&& ${\bf 200}$  &&& ${\bf 200}$ \\
$\sqrt{\beta_{N\eta'}} A_{1/2}$  &&&  $0.09\pm 0.03$ &&&  $0.09\pm 0.06$ &&& $0.06\pm 0.11$ \\
$\sqrt{\beta_{N\eta'}} A_{3/2}$  &&& $-0.16\pm 0.05$ &&& $-0.13\pm 0.09$ &&& $-0.03\pm 0.06$ \\ \hline
$P_{13}(2050)$      &&&        &&&   && \\
$M_R        $         &&& $2050\pm 4$ &&& $2045\pm 7$ &&& ${\bf 2048}$ \\
$\Gamma_R   $         &&&  $140\pm 10$ &&&  $52{+184\atop -52}$  &&& $51{+241\atop -51}$ \\
$\sqrt{\beta_{N\eta'}} A_{1/2}$ &&& $-5.71\pm 0.17$ &&& $-2.02\pm 0.26$ &&& $-3.14\pm 0.43$ \\
$\sqrt{\beta_{N\eta'}} A_{3/2}$  &&& $ 9.89\pm 0.30$ &&& $ 7.31\pm 0.93$ &&& $5.75\pm 0.79$ \\ \hline
$S_{11}(1925)$      &&&        &&&        && \\
$M_R        $         &&& $1924\pm 4$ &&& $1926\pm 10$ &&& ${\bf 1925}$ \\
$\Gamma_R   $        &&&  $112\pm 7$ &&&  $99\pm 23$  &&& $145\pm 45$ \\
$\lambda  $         &&& $1.00{+0.00\atop -0.06}$ &&& $1.00{+0.00\atop -0.98}$ &&& $1.00{+0.00\atop -0.95}$ \\
$\sqrt{\beta_{N\eta'}} A_{1/2}$  &&& $-11.84\pm 0.41$ &&& $-11.07\pm 1.43$ &&& $-19.93\pm 1.56$ \\ \hline
$P_{11}(2130)$      &&&        &&&        &&& $$ \\
$M_R        $         &&& $2129\pm 5$ &&& $2123\pm 23$ &&& ${\bf 2126}$ \\
$\Gamma_R   $         &&&  $205\pm 12$ &&& $246\pm 54$  &&& $170\pm 178$ \\
$\lambda  $         &&& $1.00 {+0.00\atop -0.04}$ &&& $1.00{+0.00\atop -0.61}$ &&& $1.00{+0.00\atop -0.95}$ \\
$\sqrt{\beta_{N\eta'}} A_{1/2}$ &&& $-11.34\pm 0.62$ &&& $-18.80\pm 0.90$ &&& $-7.45\pm 0.94$ \\
\hline\hline
\end{tabular*}
\end{center}
\end{table*}

\begin{figure*}[tb]
\includegraphics[width=1.0\columnwidth,angle=0,clip]{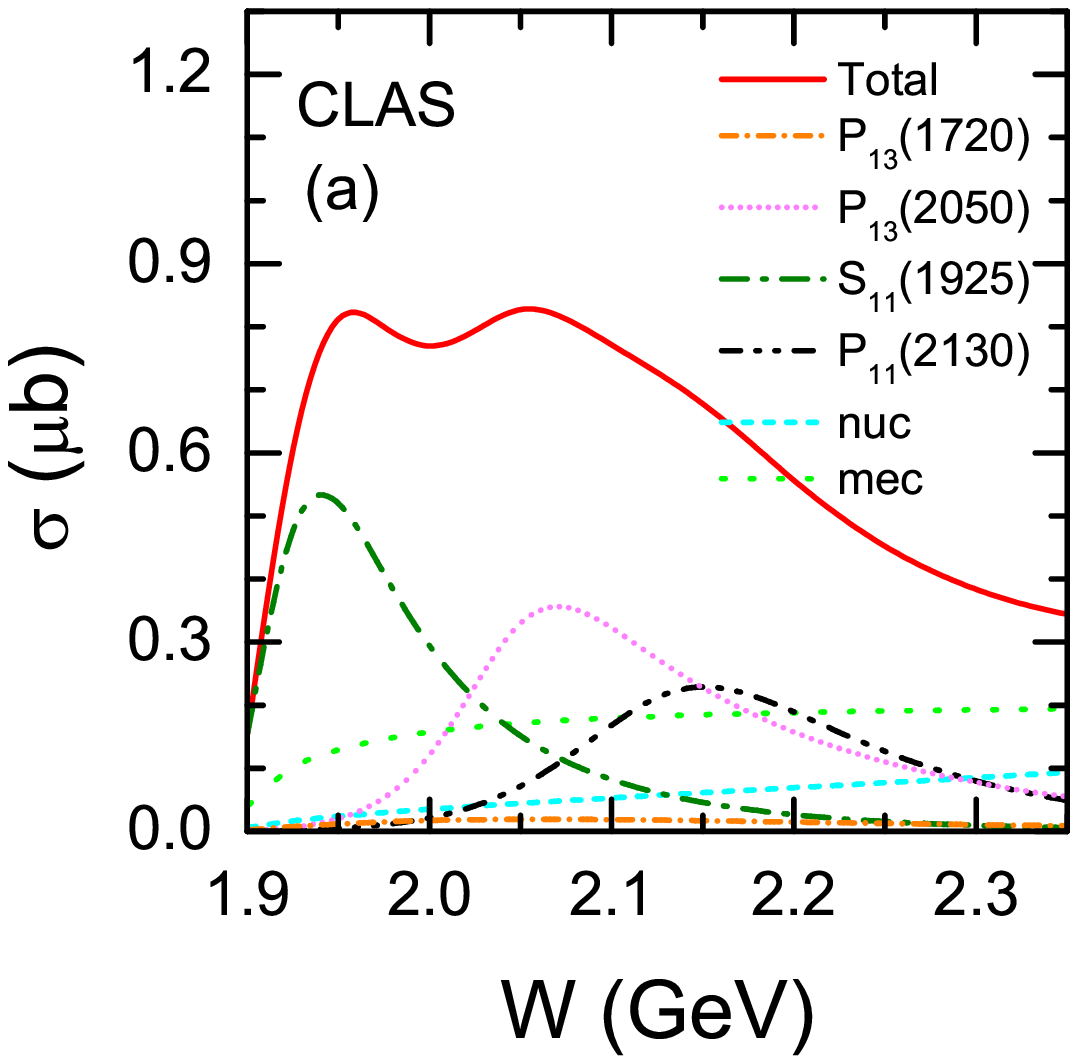}
\includegraphics[width=1.0\columnwidth,angle=0,clip]{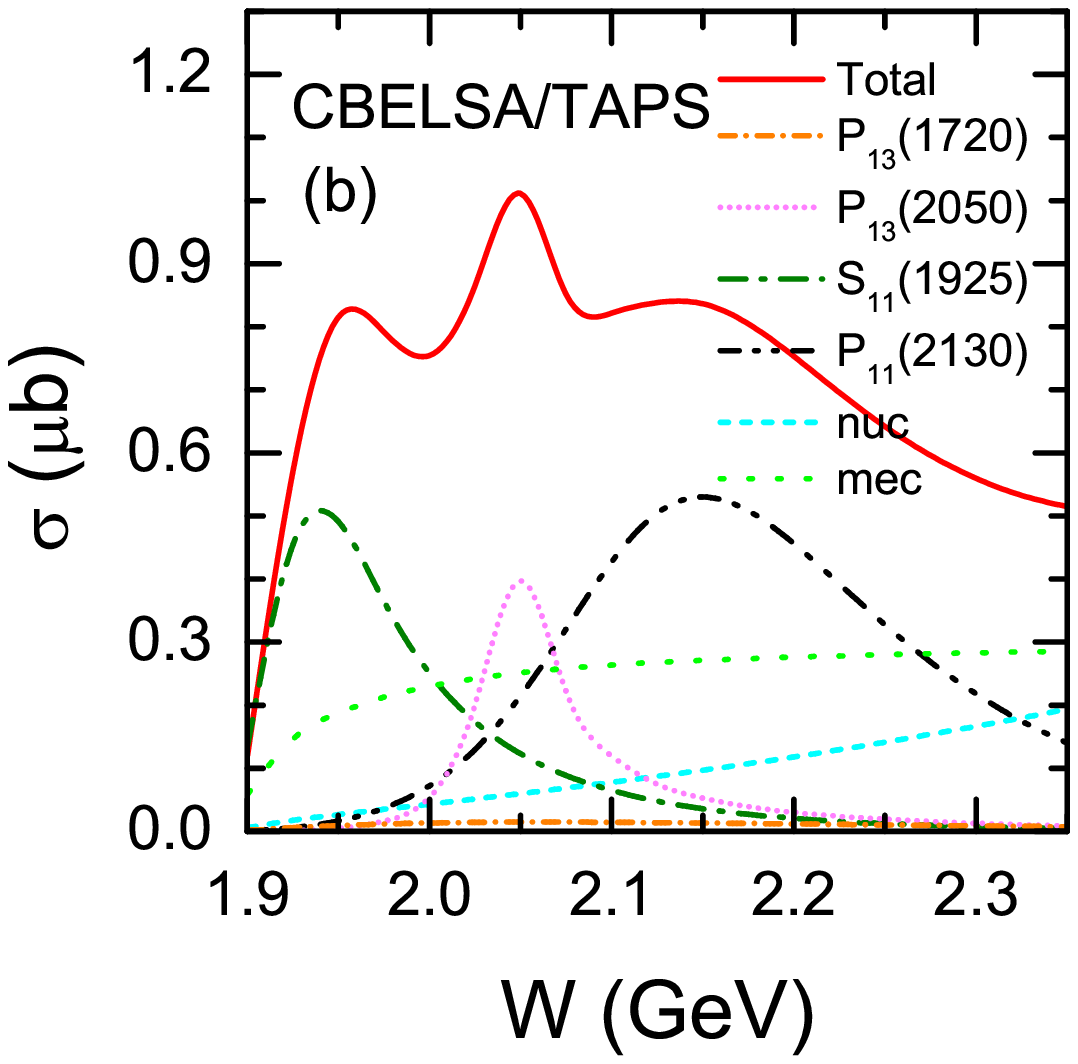}
\vskip -0.5cm
  \caption{\label{fig:4} (Color online) Total cross sections with
individual (nucleonic, mesonic, and resonance) current contributions. The left
panel (pannel (a)) results from the fit to the CLAS data while the right panel (pannel (b)) pertains to
the CBELSA/TAPS data. The corresponding parameters are given in Table
\ref{tab:1}. The peak structure at 2.05 GeV for the latter data is solely due
to the narrow $P_{13}(2050)$ resonance. }
\end{figure*}

As mentioned, Fig.~\ref{fig:3} shows the independent resulting fit curves for
the CLAS \cite{CLAS09} and for the CBELSA/TAPS \cite{CBELSA09} data. Both data
sets are reproduced with very good fit quality of $\chi^2/N =0.62$ and 0.56,
respectively.  The corresponding model parameter values are displayed in the
two columns subsumed under ``free $p$" in Table \ref{tab:1}. The uncertainties
in the resulting parameters are estimates arising from the uncertainties (error
bars) associated with the fitted experimental data points. In addition to the
resonance-mass and total-width values, the table shows the corresponding
reduced helicity amplitudes $\sqrt{\beta_{N\eta'}} A_j$, where $\beta_{N\eta'}$
denotes the branching ratio to the decay channel $N\eta'$ and $A_j$ stands for
the helicity amplitude with spin $j$. The mass and total width of the
(four-star) $P_{13}(1720)$ resonance have been fixed at the respective centroid
values quoted in PDG \cite{PDG}. Also, the fixed radiative decay branching
ratio of $\beta_{p\gamma}=0.10\%$ is well within the range of [0.05-0.25]\%
quoted in PDG for this resonance. The analysis of $\eta$ production processes
in Ref.~\cite{NOH11} yielded a value of $\beta_{p\gamma} \sim 0.12\%$. We
recall that this sub-threshold resonance is required in describing the $NN \to
NN\eta'$ reaction but not in photoproduction. As such, its contribution is
negligible here (cf. Fig.~\ref{fig:4}) and, consequently, its parameters are
not well constrained by the present photoproduction data. In particular, the
corresponding reduced helicity amplitudes quoted in Table \ref{tab:1} are
subject to much larger uncertainties than those indicated there, since the
extracted very small branching ratio $\beta_{N\eta'}$ for this resonance (cf.
Table \ref{tab:4}) depends more sensitively on the assumed mass distribution
(see discussion in Sec.~\ref{sec:NNeta'}) than the corresponding values for the
above-threshold resonances. The systematic uncertainties arising from such an
assumption are not taken into account in the error numbers given
in Table \ref{tab:1}. Following Refs.~\cite{NH04,NH06}, we also set the
radiative decay branching ratios for other resonances to be
$\beta_{p\gamma}=0.2\%$. Note that in the present tree-level calculation, the
results are rather insensitive to the branching ratios, since they enter the
model only through the total decay widths in the resonance propagator (see
Appendix \ref{sec:widths}), in addition to the fact that the results are
sensitive only to the product of the coupling constants
$g_{RN\eta'}g_{RN\gamma}$. In principle, a simultaneous consideration of
photon- and relevant hadron-induced reactions would enable us to disentangle
the individual factors contributing to this product of
coupling constants. However, the relatively poor quality of the currently
existing data for one of the relevant reactions, $\pi N \to \eta' N$ (cf.\
Fig.~\ref{fig:14} discussed in Sec.~\ref{sec:eta'N} below), does not allow
this.  We also note that the pseudoscalar-pseudovector mixing parameters
$\lambda$ (see Appendix \ref{app:hadronLagrange}) are not well constrained by
the data, as the corresponding uncertainties indicate. The fitted mass values
($M_R$) of the corresponding resonances found from the CLAS and from the
CBELSA/TAPS data are very close to each other. The same is true for all but one
of the total widths $\Gamma_R$ of these resonances at the resonance energies
$W=M_R$. A marked exception is the $P_{13}$ resonance for which the CBELSA/TAPS
data yield a much narrower width (cf.\ Table \ref{tab:1}), however, with an
associated uncertainty that is very large. We will discuss this issue in more
detail below in connection with the total cross-section results shown in
Fig.~\ref{fig:4}. The parameter values for the $S_{11}$ resonance are
practically the same for both data sets. We emphasize that the CLAS
photoproduction data \cite{CLAS09} constrain the masses of the above-threshold
resonances very well. In contrast, the CBELSA/TAPS data \cite{CBELSA09} by
themselves cannot constrain these resonance masses because of the
overdetermined fit (recall that CBELSA/TAPS data \cite{CBELSA09} can be fitted
only with one above-threshold resonance).

\begin{figure*}[tb]
\includegraphics[width=1.0\textwidth,angle=0,clip]{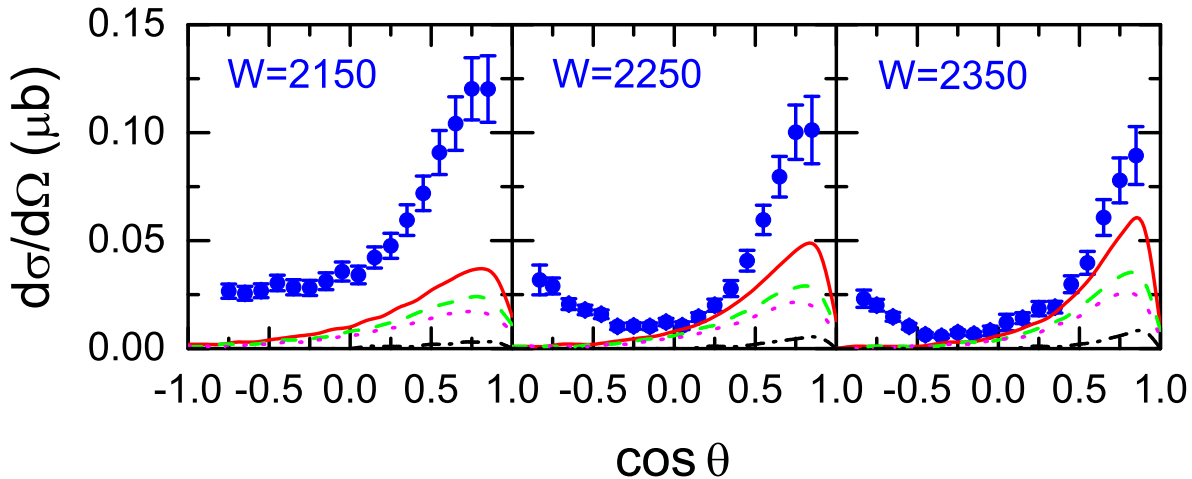}
\vskip -0.75cm
\caption{\label{fig:5}%
(Color online) Effects from $\rho$ and $\omega$ exchanges. The red solid line
is the constructive sum of the $\rho$ (green dashed)  and $\omega$ (black
dash-dotted) contributions, while the magenta dotted line is the destructive
sum of $\rho$ and $\omega$. }
\end{figure*}

\begin{figure*}[tb]
\centerline{\includegraphics[width=1.1\textwidth,angle=0,clip]{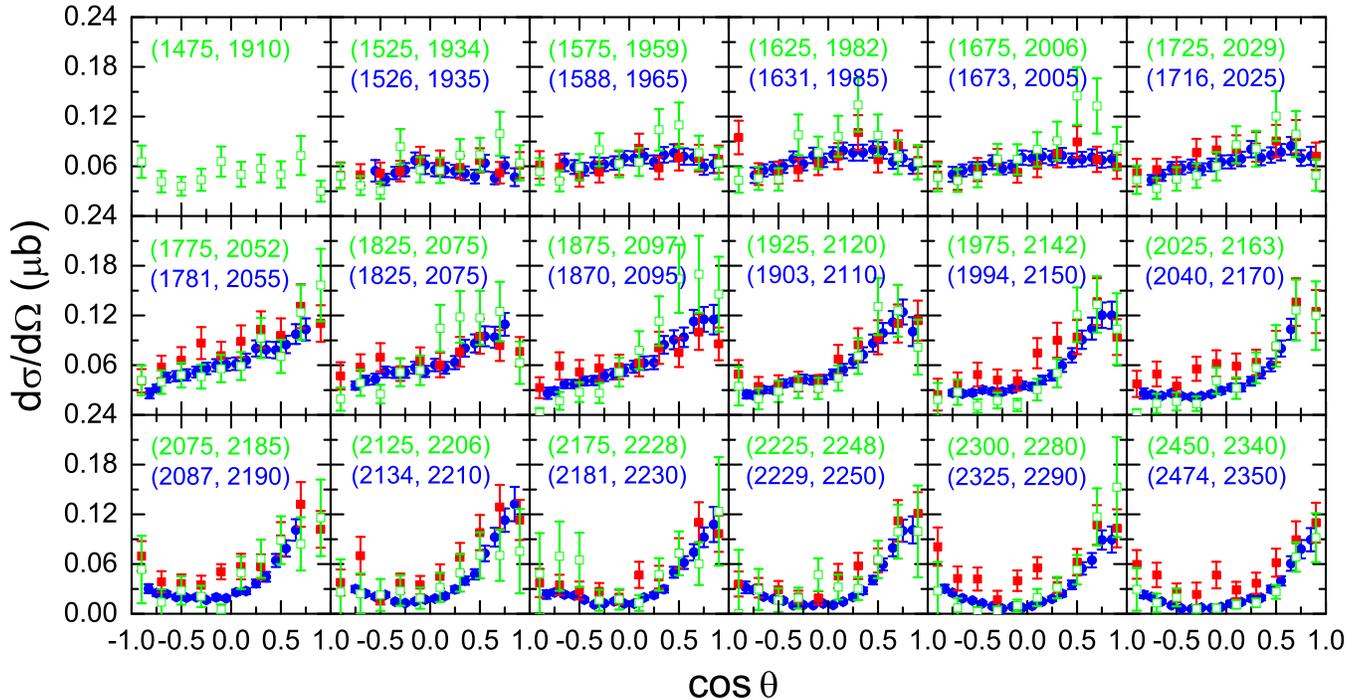}}
\vskip -0.75cm
\caption{\label{fig:6}%
 (Color online) Comparison of the quasi-free $\gamma p \to \eta' p$
differential cross-section data from CBELSA/TAPS \cite{CBELSA11} (open green
squares) with the corresponding free data from CBELSA/TAPS \cite{CBELSA09} (solid red squares) and from CLAS \cite{CLAS09} (solid blue circles).}
\end{figure*}

\begin{figure*}[tb]
\centerline{\includegraphics[width=1.1\textwidth,angle=0,clip]{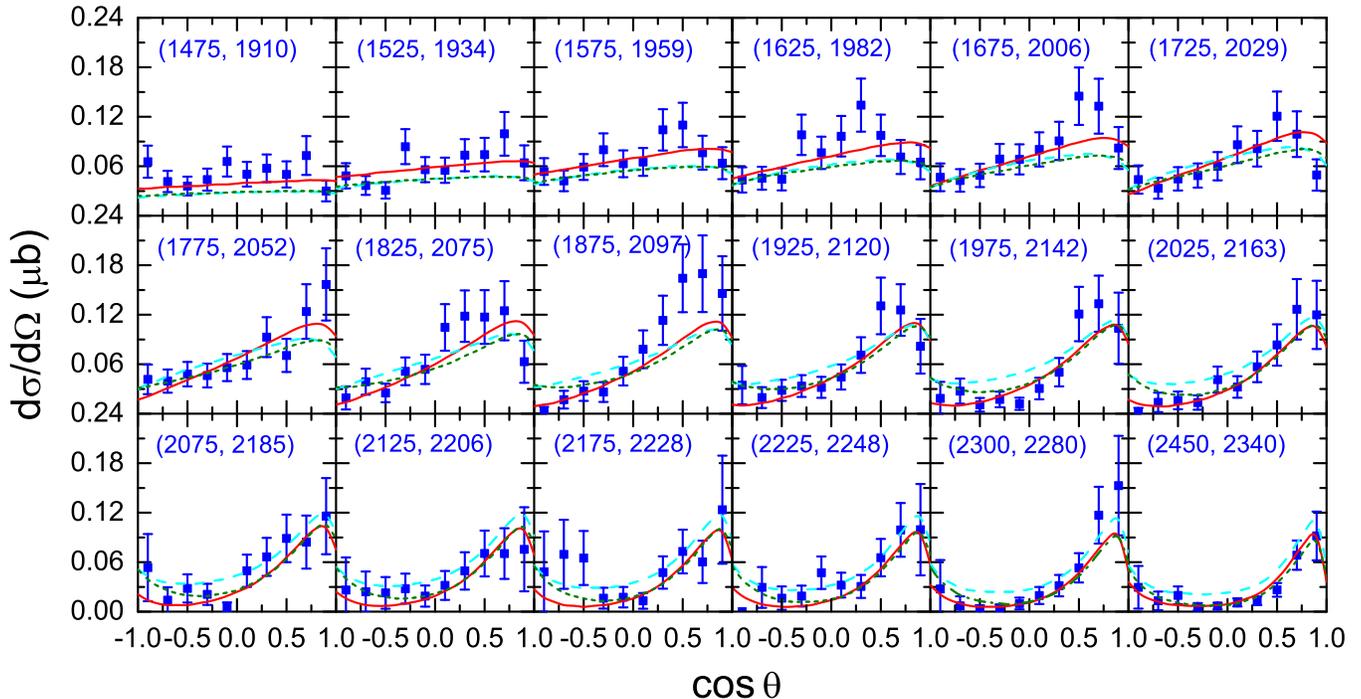}}
\vskip -0.75cm
\caption{\label{fig:7}%
(Color online) Comparison of the quasi-free $\gamma p \to   \eta^\prime p$ differential cross-section CBELSA/TAPS data of Fig.~\ref{fig:6} with theoretical results obtained via Eq.~(\ref{eq:QFXa}). The data \cite{CBELSA11}
are represented here by solid (blue) squares. The (cyan) dashed  and the (olive) short-dashed curves are obtained by respectively folding the CBELSA/TAPS and the CLAS fits of Fig.~\ref{fig:3}. The latter two folded fits have no extra parameters. The (red) solid curves provide the fit results of the quasi-free data, also obtained by using Eq.~(\ref{eq:QFXa}). The parameters of this fit are given in the right-most column of Table \ref{tab:1}.  }
\end{figure*}

\begin{figure*}[tb]
\centerline{\includegraphics[width=1.1\textwidth,angle=0,clip]{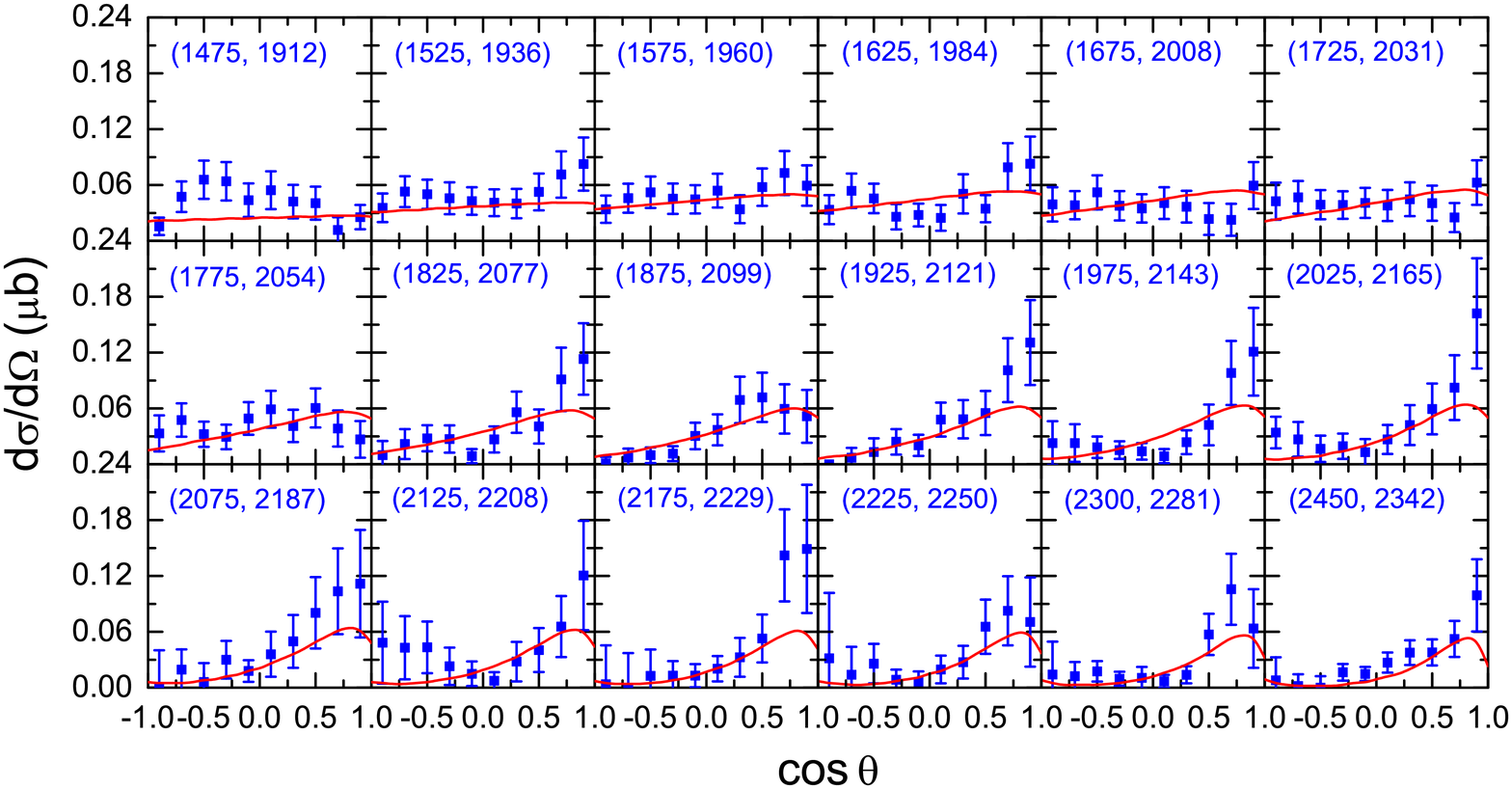}}
\vskip -0.75cm
\caption{\label{fig:8}%
(Color online) Fit results for the differential cross section of the quasi-free
$\gamma n \to \eta^\prime n$ reaction as a function of $\cos\theta$ (where
$\theta$ is the $\eta^\prime$ emission angle in the center-of-momentum frame)
for energies up to $\sqrt{s} = 2.35$ GeV.  The data are from CBELSA/TAPS \cite{CBELSA11}. }
\end{figure*}

The differences in the corresponding individual current contributions are
better seen in Fig.~\ref{fig:4}.
Comparing the resonance values found in our fits with those quoted in
PDG \cite{PDG}, our $P_{11}(2130)$, $S_{11}(1925)$, and $P_{13}(2050)$ may
perhaps be identified with the PDG resonances $P_{11}(2100)^*$,
$S_{11}(1895)^{**}$, and $P_{13}(2040)^{*}$, respectively. As mentioned in
Sec.~\ref{sec:results}, the cross section data alone in $\gamma p \to \eta' p$
cannot constrain the set of resonances uniquely. We emphasize, however, that
acceptable fits of the cross-section data for this reaction cannot be obtained
without the above-threshold $P_{13}(2050)$ resonance.

Figure \ref{fig:4} displays the individual resonance contributions to the
predicted total cross sections obtained by integrating the differential cross
section results shown in Fig.~\ref{fig:3} for the CLAS (left panel) and
CBELSA/TAPS (right panel) data. Both the nucleonic and mesonic currents yield similar
contributions for both data sets, even though they are visibly larger for the
CBELSA/TAPS data than for the CLAS data. For resonances, we see
significant differences in their relative contributions. The $P_{11}$
resonance contribution is much stronger for CBELSA/TAPS than for CLAS, which is
responsible for making the full total cross section larger for energies above
$W \sim 2.1$ GeV. For both data sets the respective $S_{11}$ resonance contributions
alone are responsible for the sharp rise of the full total cross section (red
curves) near threshold, and their corresponding parameter sets are practically
identical. By contrast, the sub-threshold $P_{13}(1720)$ resonance
contribution is negligible.

The clear peak structure in the full total cross exhibited by the CBELSA/TAPS
data is produced here by the above-threshold $P_{13}$ resonance which is much
narrower for the CBELSA/TAPS data than for CLAS.  However, the uncertainty
associated with its width is very large, as shown in Table \ref{tab:1}
and this width, therefore, is not well constrained. We recall
here that the total cross-section data were \emph{not} fitted, so the
peak structure found in the calculated results is a
consequence of the fit results of the angular distribution data as shown in
Fig.~\ref{fig:3}. The origin of this large uncertainty in the width can be
traced back to the CBELSA/TAPS measured differential cross-sections at one
particular energy only, namely at $W=2.052$ GeV, which coincides with the $P_{13}$
resonance whose position is fixed very well within our fits with very small
errors for both data sets. In view of the large error for the $P_{13}$ width
for the CBELSA/TAPS data, there is no physical significance in finding the peak
structure shown in Fig.~\ref{fig:4}. In fact, manually changing the width from
52 to 140 MeV, similar to the width found for the CLAS data, largely smoothes
out the peak resulting in an overall $\chi^2/N$ only a few percent worse than
what is reported in Table \ref{tab:1}.

The main conclusion regarding the discrepancy
between the CLAS \cite{CLAS09} and CBELSA/TAPS \cite{CBELSA09} data as
exhibited in Fig.~\ref{fig:1} is that the larger cross-section yield of the
CBELSA/TAPS data at higher energies results in a larger $P_{11}$ resonance
contribution compared to the CLAS data (compare, in
particular, the corresponding reduced helicity amplitudes in Table
\ref{tab:1}). This alone largely leads to
 the enhancement of the CBELSA/TAPS total cross section over the CLAS results
 seen in Fig.~\ref{fig:2} for energies above $W\sim 2.1$\,GeV.
For future analyses, it is important to resolve this discrepancy in the
findings of the CLAS and CBELSA/TAPS experiments if one is to obtain more
definitive answers about these above-threshold resonances.

Figure \ref{fig:5} illustrates the contributions of the $\rho$- and
$\omega$-meson exchanges to the total mesonic current. Depending on their
relative sign, one obtains constructive (red solid curves) or destructive
(magenta dotted curves) interference. The motivation for showing this detail
here is that the MAID group \cite{CYTVD03} employ $\rho$ and $\omega$ coupling
constants in their Regge contribution corresponding to a destructive
interference very close to the result shown here for a destructive
interference. By contrast, in our present calculations we have a constructive
interference between the $\rho$- and $\omega$-meson contributions following
Refs.~\cite{NH04,NH06}. Obviously, one gets different resonance parameters
whether the interference for this mesonic background is constructive or
destructive. As can be seen from Fig.~\ref{fig:5}, in the former case, the
total mesonic current (red solid curves) is relatively large at forward angles,
while in the latter case (magenta dotted curves), it is much smaller over the
entire angular range. In our model, the $\rho$ and $\omega$ coupling constants
(including the respective signs) are determined from the measured decay widths
and the SU(3) symmetry considerations in conjunction with the OZI rule (cf.\
Refs.~\cite{NH04,NH06}) which leads to a constructive interference. We were
unable to identify how the signs of the $\rho$- and $\omega$-meson
contributions were determined in Ref.~\cite{CYTVD03}.

\section{\boldmath Quasi-free $\gamma N \to \eta^\prime N$}\label{sec:quasi-free}

In this section, the quasi-free photoproduction processes $\gamma p \to \eta'
p$ and $\gamma n \to \eta' n$ are discussed. They are evaluated by folding the
corresponding cross sections for the free processes with the momentum
distribution of the target nucleon in the deuteron as described in Appendix
\ref{sec:appA} [cf. Eq.~(\ref{eq:QFXa})].

We first compare in Fig.~\ref{fig:6} the quasi-free $\gamma p \to \eta' p$
cross-section data obtained by the CBELSA/TAPS Collaboration \cite{CBELSA11}
with the corresponding free $\gamma p \to \eta' p$ data from
Refs.~\cite{CLAS09,CBELSA09}. We note that, as for the free $\gamma p \to \eta'
p$ data, the uncertainties in the quasi-free data shown here include both the
statistical and systematic errors added in quadrature. The latter errors were
quoted in Ref.~\cite{CBELSA11} but were not included in the data shown in that
reference. Here, one sees that the quasi-free data practically coincide with
the free data within their uncertainties for most of the angles and energies,
although one sees some tendency for stronger angular dependence at a few
energies. However, this does not necessarily mean that the free and quasi-free
data are compatible with each other as discussed below. It is also interesting
to note that, overall, the quasi-free CBELSA/TAPS data are more in line with
the free CLAS data \cite{CLAS09} than with the free CBELSA/TAPS data
\cite{CBELSA09}, a feature that has been also pointed out in Ref.~\cite{CBELSA11}.

If the quasi-free hypothesis holds true, one would expect that the Fermi motion of the target nucleon inside the deuteron smears out the energy dependence of the free cross section. This then should affect more the cross section at low energies where one usually observes a strong energy dependence in the corresponding free cross section (cf. Fig.~\ref{fig:2}). At higher energies, where the energy dependence of the corresponding free cross section becomes weaker, the Fermi-motion smearing should have very little effect on the cross section. Therefore, based on the free total cross section results shown in Fig.~\ref{fig:2}, it is conceivable to expect the Fermi motion to affect the cross section up to about $W \sim 2$ GeV. However, this is not what we observe in Fig.~\ref{fig:6}, where the quasi-free and free data still coincide with each other down to the lowest energy ($W=1.935$ GeV) for which the free data are available. It would be interesting to also have data for the free process for lower energies closer to the threshold energy (at 1.896 GeV) to see the energy region where Fermi motion is relevant in this reaction.

Figure \ref{fig:7} shows the same quasi-free $\gamma p \to \eta^\prime p$
differential cross-section data as in Fig.~\ref{fig:6}, but now with our
results. The cyan dashed curves are obtained by simply folding the fit results
of the free CBELSA/TAPS data \cite{CBELSA09} shown in Fig.~\ref{fig:3} with the
momentum distribution of the proton inside the deuteron according to Eq.~(\ref{eq:QFXa}), and the olive
short-dashed curves correspond to the analogous results obtained for the free
CLAS data \cite{CLAS09}. There are no extra fit parameters here; the parameters
are all pre-determined by our fits of the free $\gamma p \to \eta' p$ reaction.
First, we see that the differences between the two sets of results obtained by
folding are much smaller than what we observed in Fig.~\ref{fig:3} for the free
process at low energies. Here, the Fermi motion may be smearing out the differences observed
there. Comparing with the data, we see that the present model predictions are
overall fairly reasonable considering the fact that there are no free
parameters to fit. We reiterate here, however, that the quasi-free data exhibit
a stronger angular dependence for some energies, a finding already pointed out
in connection with the discussion of Fig.~\ref{fig:6}. Also, the present
predictions seem to exhibit a slight tendency to underestimate the lower-energy
quasi-free data. This feature will be clearer as discussed below. The
prediction corresponding to the fit result of the free CBELSA/TAPS data
\cite{CBELSA09} (cyan dashed curves) tends to overpredict the data at some higher
energies. Overall, the fits of the folded free CLAS \cite{CLAS09} and
CBELSA/TAPS \cite{CBELSA09} data have an increased $\chi^2/N$ of 1.4 and 2.5,
respectively.

 The red solid curves in Fig.~\ref{fig:7} correspond to the fit
results of the quasi-free data also obtained via Eq.~(\ref{eq:QFXa}). The resulting parameters for this fit are shown
in the right-most column of Table \ref{tab:1} for a direct comparison with
those resulting from the free data fit. Here, the resonance mass values were
fixed to be the average of the fit results of the free CLAS and CBELSA/TAPS
data, since they are well determined by these free data. As can be seen, the
fitted values of the other parameters are close to the corresponding values
obtained from the fit of the free data as expected for the quasi-free process.
The only notable difference is in the reduced helicity amplitudes,
$\sqrt{\beta_{N\eta'}} A_j$, for the spin-1/2 resonances. For the
$P_{11}(2130)$ resonance, the value is much closer to the free CLAS result,
which is understandable because at higher energies, the quasi-free data are
much closer to the free CLAS data than to the free CBELSA/TAPS data. This
might indicate a possible normalization problem in the free CBELSA/TAPS data at
higher energies. For the $S_{11}(1925)$ resonance, the reduced helicity
amplitude is almost a factor of 2 larger than the corresponding values
extracted from the free data. As can be seen in Fig.~\ref{fig:7}, this is also
easy to understand; the fit simply tries to enhance the cross section at lower
energies, where the results obtained by folding the free cross sections without
any fit parameters tend to underestimate the data.

Summarizing the discrepancies between quasi-free data and the Fermi-folded free
cross sections, we find that the folded free results tend to underestimate the
quasi-free data at lower energies and for higher energies, we find that, if
anything, the folded free results are above the quasi-free results for some
energies, in particular for the CBELSA/TAPS results. There might be two
possible causes for this energy-dependent difference between the theoretical
folding procedure and the corresponding experimental analysis:
    (\textit{i}) Our
prescription for accounting for the Fermi motion  is not quite adequate [see
Appendix \ref{app:QF}, in particular, Eq.~(\ref{eq:QFXa})]. This prescription,
however, works quite well for $\eta$ photoproduction
\cite{Krusche95,Anisovich09} where there is a much stronger energy dependence
in the cross section close to threshold. To test how sensitive the results are
on the details of the folding procedure, we have also employed the alternative
prescription of Eq.~(\ref{eq:QFXb}) which treats the total energy available to
the $\gamma p \to \eta^\prime p$ subsystem differently from Eq.~(\ref{eq:QFXa})
(for details, see Appendix \ref{app:QF}), and we found no appreciable
differences.
  (\textit{ii}) The quasi-free data contain additional nuclear
effects at lower energies which cannot be adequately described by the simple
folding procedure. In fact, our calculation shows in Fig.~\ref{fig:7} that the
effect of Fermi folding can be seen for energies up to about 2 GeV, while the
direct comparison of quasi-free and free data in Fig.~\ref{fig:6} do not show
this effect for energies down to the lowest energy of 1.934 GeV available for
the free data. We note in this context that to force the theoretical folding
results to agree with the quasi-free data at lower energies, we would need to
drastically cut the higher-momentum part of the deuteron wave function. Further
investigation are necessary to find out which of these two possible causes
applies here --- or perhaps even a combination of both.

One purpose of fitting the quasi-free data here is to have a quasi-free proton
result on the same footing as the quasi-free neutron result, since the latter
has to be fitted to the corresponding quasi-free data to fix the resonance
transition electromagnetic couplings. In this way, quantities such as the ratio
of the neutron and proton branching ratios will be free of possible unwanted
effects which may distort the results otherwise.

In Fig.~\ref{fig:8} we show the results for the quasi-free $\gamma n \to \eta'
n$ reaction. As mentioned in Sec.~\ref{sec:results},
incorporating this reaction in our combined analysis rules out
the set of above-threshold resonances $\{P_{11},P_{13},D_{13}\}$ which
otherwise for the free $\gamma p \to \eta' p$ reaction  reproduces the data
just as well as the set $\{S_{11},P_{11},P_{13}\}$. Overall, we reproduce the
quasi-free $\gamma n \to \eta' n$ data reasonably well, with $\chi^2/N =
0.82$. The calculation requires adjusting free resonance-neutron-$\gamma$
($Rn\gamma$) coupling constants that need to be determined through the combined
fit of the photon- and hadron-induced reaction data. Table \ref{tab:3} displays
the resulting values. For the $P_{13}(1720)$ resonance, we obtain
$\beta_{n\gamma}=0.016\%$ which is at the upper limit of the range of [0.0 -
0.016]\% quoted in PDG \cite{PDG}. We note that, as in the free $\gamma p \to
\eta' p$ case, here the parameters associated with this sub-threshold resonance
are not well constrained by the data. In Table \ref{tab:3}, we also give the
ratio of the neutron-to-proton branching ratios. The sign of this ratio
reflects the relative sign in the corresponding neutron and proton
electromagnetic coupling constants. Although the branching ratios may be
subject to a considerable ambiguity since, as discussed in the previous
section, only the product of the coupling constants $g_{RN\eta'}g_{RN\gamma}$
is well determined in the present calculation, the ratio
$\beta_{n\gamma}/\beta_{p\gamma}$ is free of such an ambiguity.

\begin{table}[tb]
\caption{\label{tab:3}
Electromagnetic couplings extracted from the CBELSA/TAPS quasi-free neutron
data \cite{CBELSA11} in a global fit of the photon- and hadron-induced
reactions data. The corresponding branching ratios of resonances decaying into
$n\gamma$ and $p\gamma$, $\beta_{n\gamma}/\beta_{p\gamma}$, are listed in the
last row. The sign of this ratio reflects the relative sign in the
corresponding neutron and proton electromagnetic coupling constants.}
\begin{center}
\begin{tabular*}{\columnwidth}{@{\extracolsep\fill}lrrrr}
\hline\hline
           & $P_{13}(1720)$ & $P_{13}(2050)$ & $S_{11}(1925)$ & $P_{11}(2130)$  \\ \hline
$\sqrt{\beta_{N\eta'}} A_{1/2}$  & $0.04$ & $0.94$ & $15.54$ & $7.60$ \\
$\sqrt{\beta_{N\eta'}} A_{3/2}$  & $-0.00$ & $-1.64$  & --- &  --- \\ \hline
$\beta_{n\gamma}/\beta_{p\gamma}$  &  $0.32$        &     $-0.09$     &      $-0.61$    &     $-3.06$ \\
\hline\hline
\end{tabular*}
\end{center}
\end{table}

\begin{table*}[tb]
\caption{\label{tab:4} The (hadronic) parameter values as determined from the
combined fit to the $\gamma N \to \eta^\prime N$ (parameter values given in
Table \ref{tab:1}, free $p$ CLAS), $\pi N \to \eta^\prime N$, and $NN \to
NN\eta^\prime$ reaction data.
 The values in boldface were kept fixed during the fit procedure.
The values in square brackets for the branching ratios are the PDG quotes.
$(g_{NN\eta^\prime}, \lambda) = (1.00, 0.53)$. }
\begin{center}
\begin{tabular*}{\textwidth}{@{\extracolsep\fill}lrrrr}
\hline\hline
parameters                             & $S_{11}(1925) $ & $P_{11}(2130) $ & $P_{13}(1720) $ & $P_{13}(2050) $ \\
\hline
$M_{R}^{}$ (
MeV)                  &   1924     &     2129    &   {\bf 1720}   &   2050    \\
$\Gamma_{R} $ (MeV)          &     112     &      205     &     {\bf 200}   &     140    \\
\hline
$\beta_{N\eta'}^{}$ (\%)         &    6    &     3   &    0.09   &    2    \\
$\beta_{N\pi}^{}$ (\%)            &  22    &    25  &   [$11 \pm 3$]   16    &    25    \\
$\beta_{N\eta}^{}$ (\%)          &   4     &  [$61\pm 60$]  0.5   &   [$4.0\pm 1.0$] 9   &   0.03    \\
$\beta_{N\rho}^{}$ (\%)         &   22   &    62   &   [70-85] 75   &    37   \\
$\beta_{N\omega}^{}$ (\%)   &   47   &    13   &     2    &  36   \\
\hline \\
($g_{RN\eta'}^{}$, $\lambda$) & (0.68, 1.00)  & (1.77, 1.00) & (1.20,  ---) & (1.38,  ---)   \\
($g_{RN\pi}^{}$, $\lambda$)    & (-0.36, 1.00) & (-1.28, 1.00) & (-0.17,  ---) & (-0.12,  ---)    \\
($g_{RN\eta}^{}$, $\lambda$) &(-0.28, 0.81)  & (-0.35, 0.34) & (-1.50, ---) & (-0.04., ---) \\
($g^{(1)}_{RN\rho}, g^{(2)}_{RN\rho}, g^{(3)}_{RN\rho}$)  &
(-2.42, 0.04, ---) & (2.58, -0.14, ---) & (-23.63, 54.09, 16.72) & (0.50, 9.10, 28.66) \\
($g^{(1)}_{RN\omega}, g^{(2)}_{RN\omega}, g^{(3)}_{RN\omega}$)  &
(1.02, -1.70, ---) & (2.47, 0.53, ---) & (-27.64, 138.87, -318.85) & (-3.19, -16.75, -36.39) \\
\hline\hline
\end{tabular*}
\end{center}
\end{table*}

\begin{figure}[tb]
\includegraphics[width=1.0\columnwidth,angle=0,clip]{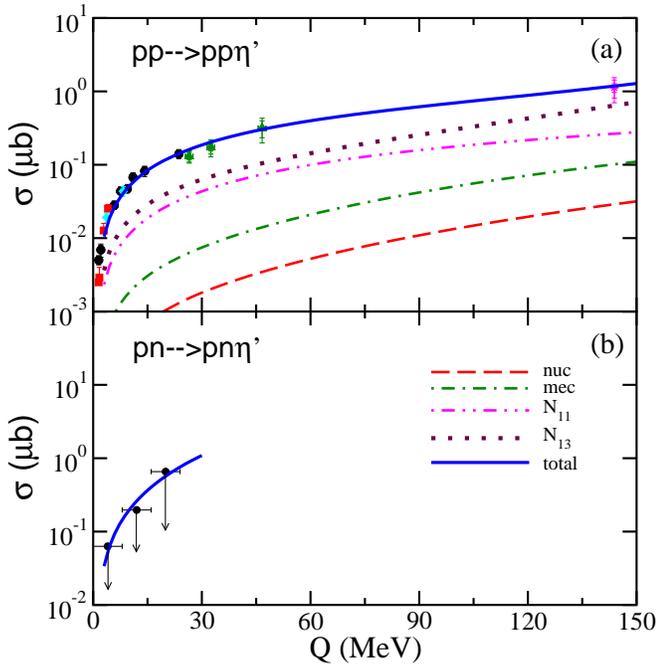}
\caption{\label{fig:10}
(Color online) Total cross sections for $pp \to pp\eta^\prime$ (panel (a)) and  
$pn \to pn\eta^\prime$ (panel (b)) as functions of
the excess energy $Q\equiv\sqrt{s}-\sqrt{s_0}$, where
$\sqrt{s_0}=2m_N+m_{\eta^\prime}$. The results (blue solid curves) correspond
to the parameter set determined in conjunction
with the fit to the $\gamma N \to \eta' N$ and $\pi N \to \eta' N$ data. For
$pp\eta'$, the individual current contributions are also show: nucleonic
current (red dashed curve), mesonic current (green dash-dotted),
$N_{11}=S_{11}(1925)+P_{11}(2130)$ resonance current (magenta dash-double-dotted),
$N_{13}=P_{13}(1720)+P_{13}(2050)$ (maroon dotted).  The $pp\eta'$ data are from Refs.~\cite{ppetapdata,Balestra,Khoukaz}; the $pn\eta'$ data, which are upper limits, are from Ref.~\cite{COSY11-10b}.}
\end{figure}

\begin{figure*}[tb]
\includegraphics[width=0.8\textwidth,angle=0,clip]{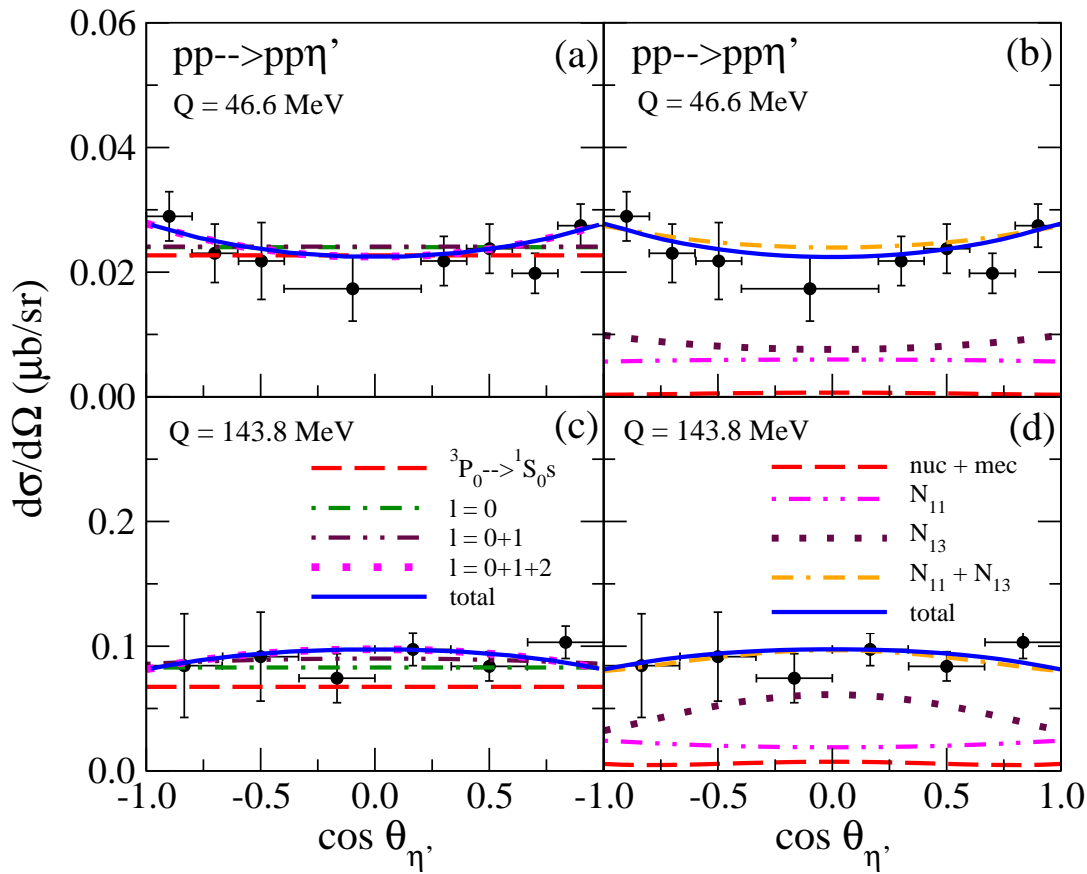}
\caption{\label{fig:11}
(Color online) $\eta^\prime$ angular distribution in $pp \to pp\eta^\prime$ in
the center-of-momentum frame of the system for two excess energies of $Q=46.6$
(panels (a) and (b)) and 143.8 MeV (panels (c) and (d)).  
The results (blue solid curves) correspond to the parameter set
determined in conjunction with the fit to
the $\eta'$ photoproduction. The panels (a) and (c) in left hand-side column show the
contributions from the partial waves ${{}^3P_0} \to {{}^1S_0s}$ (red dashed
curve), sum of the $s$-wave ($l=0$) $\eta'$ (green dash-dotted), $s+p\,$-waves
($l=0,1$) (maroon dash-double-dotted), and $s+p+d\,$-waves ($l=0,1,2$) (magenta
dotted). The panels (b) and (d) in the right-hand-side column show the nucleonic+mesonic
currents (red dashed), the spin-1/2 resonances (green dash-dotted), the
spin-3/2 resonances (maroon dash-double-dotted), and the sum of the spin-1/2
and -3/2 resonances (magenta dotted) contributions. The data are from the
COSY-11 collaboration (46.6 MeV) \cite{Khoukaz} and from DISTO  (143.8
MeV) \cite{Balestra}. }
\end{figure*}

\begin{figure}[tb]
\includegraphics[width=1.0\columnwidth,angle=0,clip]{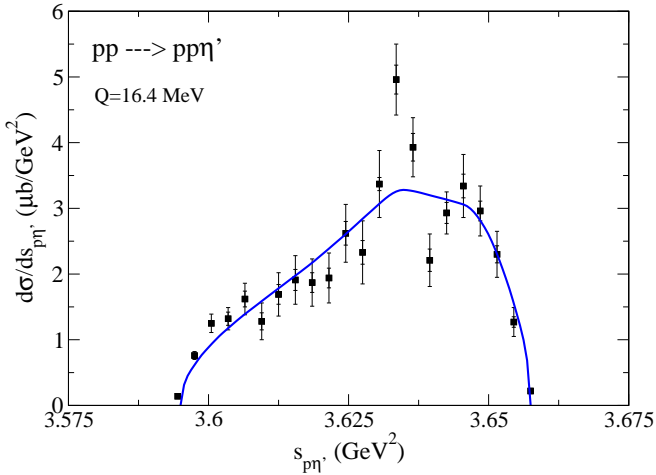}
\caption{\label{fig:12}
(Color online) The $p\eta^\prime$ invariant mass distributions in $pp \to
pp\eta^\prime$ at an excess energy of $Q=16.4$ MeV. The blue solid
curve corresponds to the full result
determined in conjunction with the fit to the $\eta'$ photoproduction data. The
data are from Ref.~\cite{COSY11-10a}. Here, the theoretical results have been
multiplied by an arbitrary normalization factor of 1.478 in order to facilitate
the comparison of the shape with the data.}
\end{figure}

\begin{figure*}[tb]
\includegraphics[width=0.9\textwidth,angle=0,clip]{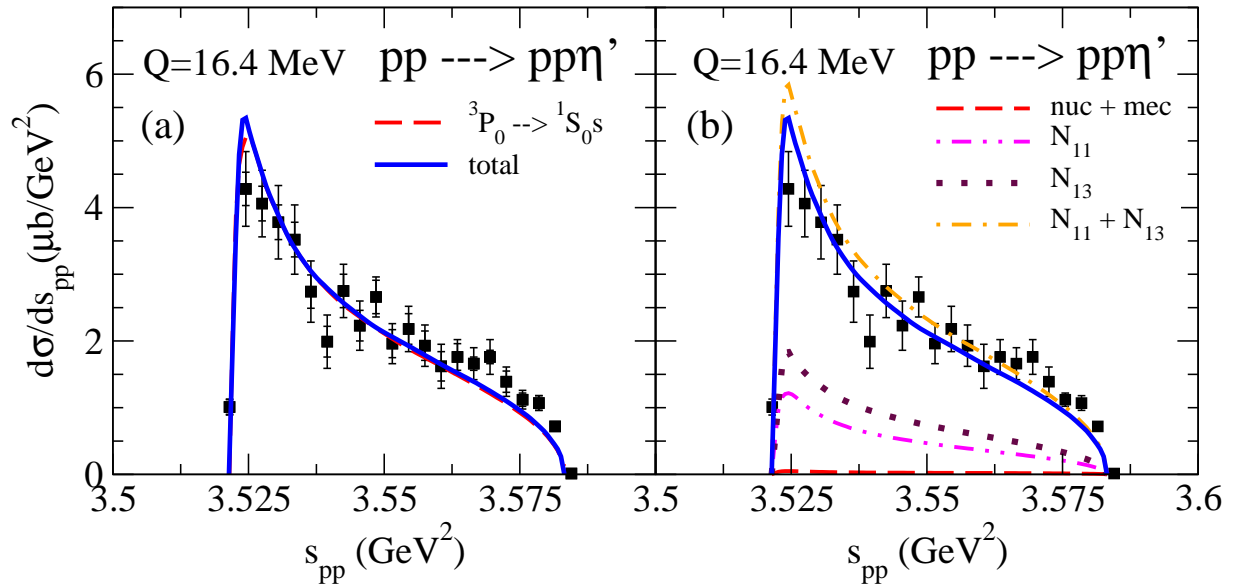}
\caption{\label{fig:13}
(Color online) The $pp$ invariant mass distributions in $pp \to pp\eta^\prime$
at an excess energy of $Q=16.4$ MeV. The blue solid curves correspond
to the full result determined in conjunction
with the fit to the $\eta'$ photoproduction data. The left panel (a) shows the
$^3P_0 \to ^1S_0s$ (red dashed curve) partial wave contributions. The right
panel (b) shows the nucleonic+mesonic currents (red dashed), the spin-1/2
resonances (green dash-dotted), the spin-3/2 resonances (maroon
dash-double-dotted), and the sum of the spin-1/2 and -3/2 resonances (magenta
dotted) contributions. The data are from Ref.~\cite{COSY11-10a}. Here, the
theoretical results have been multiplied by an arbitrary normalization factor
of 1.478 in order to facilitate the comparison of the shape with the data.}
\end{figure*}

\begin{figure*}[tb]
\vglue 0.5cm
\includegraphics[width=0.95\columnwidth,angle=0,clip]{dmpp_S11P11P13P13-1_Nps_new_03_466.eps}
\includegraphics[width=0.95\columnwidth,angle=0,clip]{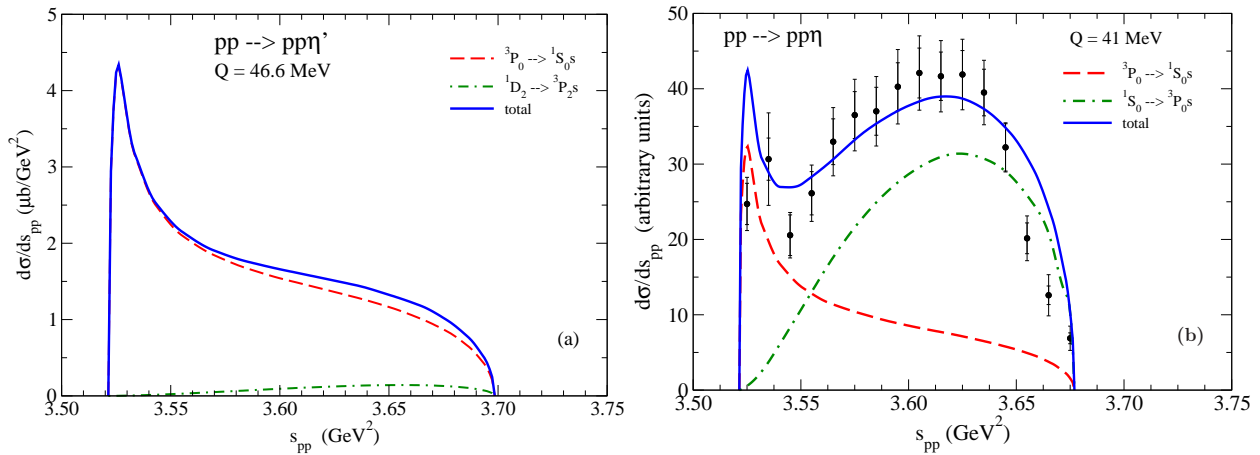}
\caption{\label{fig:14}
(Color online) Left panel (a): prediction for the $pp$ invariant mass distribution in $pp \to
pp\eta^\prime$ at an excess energy of $Q=46.6$ MeV. The blue solid
curve corresponds to the result determined in conjunction with the fit to the $\eta'$ photoproduction data. The
${}^3P_0 \to {}^1S_0s$ (red dashed curve) and the ${}^1D_2 \to {}^3P_2s$ (green
dash-dotted curve) partial wave contributions are also shown.
Right panel (b): the model result (blue solid curve) of Ref.~\cite{NHHS03} for the $pp$ invariant mass distribution in $pp \to
pp\eta$ at an excess energy of $Q=41$ MeV. The two dominant partial-wave contributions,
${}^3P_0 \to {}^1S_0s$ (red dashed curve) and ${}^1S_0 \to {}^3P_0s$ (green dash-dotted curve), are also  shown. The data are from Ref.~\cite{COSYTOF03}. }
\end{figure*}

\begin{figure}[tb]
\includegraphics[width=1.0\columnwidth,angle=0,clip]{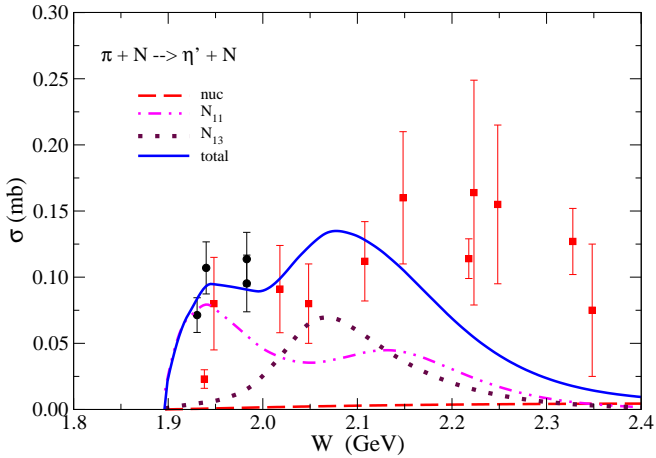}
\caption{\label{fig:9}
(Color online) Total cross section for $\pi N \to \eta^\prime N$ as a function
of the total energy of the system. The blue solid curves
corresponds to the full results, determined in
conjunction with the fit to the $\gamma N \to \eta' N$ as well as the $NN \to
NN\eta'$ data. The individual current contributions are also show: nucleonic
current (red dashed curve), $N_{11} = S_{11}(1925) + P_{11}(2130)$ resonance
current (magenta dashed-double-dotted), and $N_{13} = P_{13}(1720)+P_{13}(2050)$
 (maroon dotted). Data for $\pi^- p \to \eta' n$
(black solid circle) and for $\pi^+ n \to \eta' p$ (red solid square) are from
Refs.~\cite{Dufey68,Rader72,Baldini88}.}
\end{figure}

\section{\boldmath $NN \to NN \eta^\prime$}\label{sec:NNeta'}

In this section, we discuss our results for the nucleon-induced reaction $NN
\to NN\eta'$ obtained from the combined analysis of this reaction
together with the $\pi N \to \eta' N$ and $\gamma N \to \eta' N$ reactions. As
mentioned in Sec.~\ref{sec:results}, we find that among the spin-1/2 and -3/2
resonances considered in this work, the set of above-threshold resonances
$\{S_{11},P_{11},P_{13}\}$ plus the sub-threshold $P_{13}(1720)$ resonance
yields the best fit to all the available $\eta'$ production data in the energy
range  considered in this work for photon- and hadron-induced reactions. The
sub-threshold $P_{13}(1720)$ resonance, in particular, is required to help
reproduce the observed shape of the $\eta'$ angular distribution at the excess
energy of $Q=46.6$ MeV measured by the COSY-11 Collaboration \cite{Khoukaz}. We
recall here that in the present analysis, the mass of the sub-threshold
$P_{13}$ as well as its width were fixed at the outset at the centroid values
of the $P_{13}(1720)^{****}$ quoted in PDG \cite{PDG}. As has been mentioned in
Sec.~\ref{sec:results}, we also note that the $NN \to NN\eta'$ reaction rules
out the set of above-threshold resonances $\{S_{11},P_{11},D_{13}\}$ which fits
the photoproduction data as well as the set $\{S_{11},P_{11},P_{13}\}$. As
discussed later in connection with the invariant $pp$ mass
distribution in Fig.~\ref{fig:13}, the set $\{S_{11},P_{11},D_{13}\}$ is unable
to reproduce the measured invariant $pp$ mass distribution. In the following,
for simplicity, we restrict ourselves to the results obtained in conjunction
with the CLAS photoproduction data (cf. Table \ref{tab:1}, column labeled
``free $p$" CLAS) because the results corresponding to the other parameter sets
shown in Table \ref{tab:1} can fit the $NN\eta'$ data equally well.

The set of the parameter values of the present model that directly affect the
hadronic processes is displayed in Table \ref{tab:4}. The values result from
the combined fit of the $\eta'$ production in photon- and hadron-induced
reactions. The resonance partial decay widths were calculated by folding the
partial decay widths for a given decaying resonance mass and a given emitted
meson mass with the corresponding mass distributions. For the latter, we assume
Gaussian distributions with widths given by the corresponding total widths of
the resonance and of the meson. The branching ratio $\beta_{N\eta'}$ for the
sub-threshold $P_{13}(1720)$ resonance is extremely small because this
arises only from the far (upper) tail of its mass distribution. As such, it is subject
to considerable uncertainties. In the present work, the resonance
coupling constants cannot be determined uniquely because the available data for
$\eta'$ production are not sufficient to impose more stringent constraints. In
particular, the lack of $pn \to pn \eta'$ data
--- there exist only three upper-limit total cross-section data points (see
Fig.~\ref{fig:10}) --- makes it difficult to constrain quantitatively the
relative contributions of the isoscalar ($\eta$, $\omega$) and isovector
($\pi$, $\rho$) meson exchanges. It is also clear that one needs to consider
meson-production reactions other than $\eta'$ production to better constrain
the relevant coupling constants.

Our result for the $pp \to pp\eta'$ total cross section is shown in
Fig.~\ref{fig:10}. We see that the data are nicely reproduced over a wide range
of excess energy. The dynamical content of the present model is also displayed.
One sees that the spin-1/2 and -3/2 resonance contributions [in Fig.~\ref{fig:10},
$N_{11}=S_{11}(1925)+P_{11}(2130)$ and $N_{13}=P_{13}(1720)+P_{13}(2050)$,
respectively] have different energy dependencies. In the lower
excess-energy region, the spin-1/2 resonance contribution is only slightly
smaller than the spin-3/2 resonance contribution, but as the energy increases,
the spin-3/2 resonance contribution starts to dominate. Here, the dominant spin-3/2
resonance contribution is due to the $P_{13}(2050)$, while the dominant
spin-1/2 resonance contribution is from the $S_{11}(1925)$ resonance. Although
the overall relative $N_{11}$ and $N_{13}$ resonance-set contributions are well
determined, the individual spin-1/2 and -3/2 resonance contributions within
the set are not well constrained by the existing data, which is the reason why
we only show the contributions of the sums of the spin-1/2 ($N_{11}$) and the
spin-3/2 ($N_{13}$) resonances. The nucleonic and mesonic currents yield
contributions that are much smaller than those of the resonances in the entire
excess-energy range shown in Fig.~\ref{fig:10}. The result for the $pn \to
pn\eta'$ total cross section is also shown in Fig.~\ref{fig:10}. The $pn\eta'$
together with the $pp\eta'$ reactions helps constraining the isoscalar- and
isovector-meson couplings (here, $M=\pi, \eta, \rho, \omega$) to the
resonances. Unfortunately, only the upper limit of the cross section in a
limited energy range is currently available for the $pn\eta'$ process.
Therefore, the isoscalar-isovector meson exchange content of the present model
is subject to this limitation in the existing data.

The results for the $\eta'$ angular distribution in $pp \to pp\eta'$ at the
excess energies of $Q=46.6$ and 143.8 MeV are shown in Fig.~\ref{fig:11}. The
data are reproduced very well. We recall that the sub-threshold $P_{13}(1720)$ resonance
is needed to reproduce the experimentally observed $\eta'$ angular distribution,
especially, at $Q=46.6$ MeV.
Here, one might regard this resonance as simulating some missing background in the
present model. In this connection, however, we mention that this resonance is the closest
known resonance to threshold that helps to reproduce the measured angular distribution.
In the left-hand-side column, at $Q=46.6$ MeV,
we see that the angular distribution is dominated by the $\eta'$ in the
$s$-wave due to the completely dominant transition  ${{}^3P_0} \to {{}^1S_0s}$.
At the higher energy of $Q=143.8$ MeV, the $s$-wave contribution still
dominates to a large extent, leading to a nearly flat angular distribution as
exhibited by the data, even though the ${{}^3P_0} \to {{}^1S_0s}$ partial-wave
contribution is somewhat smaller at this high energy than at the lower energy
of $Q=46.6$ MeV. As can be seen in the panels on the right-hand-side column in
Fig.~\ref{fig:11}, the flat angular distribution, especially at higher energy,
is achieved in the present model by an interference among different currents,
in particular, between the spin-1/2 and -3/2 resonance currents.

In Fig.~\ref{fig:12}, the result for the $p\eta'$ invariant mass distributions
at an excess energy of $Q=16.4$ MeV is shown. The data are reproduced well.

The $pp$ invariant mass distribution together with the $\eta'$ angular
distribution poses a relatively strict constraint on the set of resonances,
provided they are above-threshold resonances.  In particular, the set of
above-threshold resonances $\{S_{11},P_{11},D_{13}\}$, which describes the
photoproduction data as well as the set $\{S_{11},P_{11},P_{13}\}$, is unable to
reproduce the higher energy region of the measured $pp$ invariant mass
distribution by the COSY-11 collaboration \cite{COSY11-10a}.

As we shall discuss below, the present result for the $pp$ invariant mass
distribution in $pp \to pp\eta'$ reaction has an interesting implication on the
issue of the reaction mechanisms in the $pp \to pp\eta$ reaction. In the latter
reaction, there has been observed a significant enhancement of the cross
section for larger $pp$ invariant mass values compared to that given by the
phase-space plus the $pp$ FSI \cite{COSYTOF03,COSY1104}.  One of the possible
explanations for this enhancement is the relatively strong $N\eta$ FSI.
However, as pointed out in Ref.~\cite{COSY11-10a}, the shape of the $pp$
invariant mass distribution data in $pp\eta$ at $Q=15.5$ MeV is practically the
same as that of $pp\eta'$ shown in Fig.~\ref{fig:13}.
Since the $N\eta'$ FSI is much smaller than the $N\eta$ FSI, the explanation of
the enhancement based on the $N\eta$ FSI was ruled out in $pp \to pp\eta$ in
Ref.~\cite{COSY11-10a}. In Ref.~\cite{NHHS03}, an alternative explanation,
based on the higher partial-wave (final state $P$-wave) contribution, was
proposed, together with a way to verify this proposed mechanism in a
model-independent manner. Yet another explanation is the energy dependence of
the basic production amplitude $J$ [cf. Eq.~(\ref{eq:C1})] as proposed in
Ref.~\cite{Delof}.

Figure \ref{fig:13} shows our results for the $pp$ invariant mass distribution
in $pp \to pp\eta'$ revealing a good agreement with the COSY-11 data
\cite{COSY11-10a}. In the left panel of Fig.~\ref{fig:13} one sees that the
$pp$ invariant mass distribution is practically exhausted by the ${{}^3P_0} \to
{{}^1S_0s}$ partial wave. This is quite surprising in view of the findings of
Ref.~\cite{NHHS03} mentioned above, where a significant final-state $P$-wave
contribution was found in the higher $pp$ invariant mass region in the $pp \to
pp\eta$ reaction. The present finding implies that the $S$-wave basic
production amplitude ($J$) in the present model should have an energy
dependence as proposed in Ref.~\cite{Delof}, since the $pp$ invariant-mass
dependence introduced by the $pp$ FSI is not enough to account for the
enhancement of the measured $pp$ invariant-mass distribution at larger
invariant masses. This finding tells us that the conclusion reached in
Ref.~\cite{COSY11-10a} ruling out the $N\eta$ FSI as a possible source of the
enhancement in the $pp$ invariant mass distribution at larger invariant mass
values based on the comparison of the corresponding shapes in $pp\eta$ and
$pp\eta'$ has to be taken with caution since there might be different
mechanisms operating in these reactions as shown explicitly here in
Fig.~\ref{fig:13}. At this stage, it is natural to ask what the
underlying dynamics is in the $S$-wave contribution that accounts for
the enhancement of the $pp$ invariant mass distribution at larger invariant
mass values in $pp \to pp\eta'$ as compared to that in the $pp \to pp\eta$
reaction, where the enhancement arises from the ${}^1S_0 \to {}^3P_0s$ partial
wave. In the right panel of Fig.~\ref{fig:13}, we show the individual current
contribution to the $pp$ invariant mass distribution. We see that the
enhancement at higher values of invariant mass is largely due to the
constructive interference between the spin-1/2 (green dash-dotted curve) and
the spin-3/2 (maroon dash-double-dotted curve) resonance contributions. The
present model prediction for $pp$ invariant-mass distribution at a higher
excess energy of $Q=46.6$ MeV is shown in the left panel of Fig.~\ref{fig:14}.
Here one sees an onset of the $^1D_2 \to {{}^3P}_2s$ contribution. This result
together with the result in Fig.~\ref{fig:13} reveal that the $P$-wave
contribution in $pp \to pp\eta'$ is much smaller than that in $pp \to pp\eta$.
For a close comparison with the $pp$ invariant mass distribution in the $pp \to
pp\eta$ reaction at higher excess energies, we show in the right panel of
Fig.~\ref{fig:14} the corresponding results at $Q=41$ MeV from
Ref.~\cite{NHHS03}.  Here we see a striking difference between the model
results for the two reactions. In $pp \to pp\eta$, there is a very large
enhancement in the $pp$ invariant mass distribution at higher invariant mass
values due to the ${}^1S_0 \to {}^3P_0s$ partial wave contribution.
By contrast, this partial wave contribution is minimal in the $pp \to
pp\eta'$ reaction. It would be very interesting to have measurements of the
$pp$ invariant mass distribution at this excess energy in $pp \to pp\eta'$ to
verify the present model prediction.
 Obviously, these are model-dependent results. As such, it will be very
 interesting to verify them in
a model-independent manner as pointed out in Ref.~\cite{NHHS03}.

\section{\boldmath $\pi N \to \eta^\prime N$}\label{sec:eta'N}

Experimental data for the $\pi N \to \eta' N$ reaction are scarce. The only
available data are the total cross sections for $\pi^- p \to \eta' n$ and
$\pi^+ n \to \eta' p$ \cite{Dufey68,Rader72,Baldini88} which are subject to
large uncertainties, as can be seen in Fig.~\ref{fig:9}. Notwithstanding the
fact that these data offer little constraints for the model parameters, they
were included in the global fit and the corresponding fit results are also
shown in Fig.~\ref{fig:9}. We note here that within the present model the
results for $\pi^- p \to \eta' n$ and $\pi^+ n \to \eta' p$ are identical. An
interesting feature of the present model result is the double-bump structure
caused by the $S_{11}(1925)$ and an interplay of the $P_{13}(2050)$ and 
$P_{11}(2130)$ resonances (cf.\ Table \ref{tab:4}). The $S_{11}(1925)$ 
resonance is just about 20 MeV above
threshold. In view of their large uncertainties, the currently existing data
shown in Fig.~\ref{fig:9} can indeed accommodate such a structure, however,
clearly more accurate data are needed for a definitive answer. If
experimentally corroborated, such a bump structure would rule out the
sub-threshold resonance-dominance assumption of Ref.~\cite{Cao08}, where
$S_{11}(1535)$ resonance dominance is assumed to describe both the $\pi N \to
\eta' N$ and $NN \to NN\eta'$ cross-sections data since it is not possible to
generate any bump structure from sub-threshold resonances alone.

\section{Summary}\label{sec:conclusions}

In the present work, we have revisited the theoretical description of $\eta'$
production in photon- and nucleon-induced reactions to take into account the
recent additions of accurate data to the corresponding world data base
\cite{CLAS09,CBELSA09,CBELSA11,COSY11-10a,COSY11-10b}. All of the currently
available data in the resonance-energy region considered in this work are
nicely reproduced within the present  model in a combined analysis of the
reactions $\gamma N \to N \eta'$, $NN \to NN\eta'$, and $\pi N \to\eta'N$.
Considering only spin-1/2 and -3/2 resonances, we have found that the data are
reproduced with a minimum of four resonances, i.e. $P_{13}(1720)$,
$S_{11}(1925)$, $P_{11}(2130)$, and $P_{13}(2050)$. The $P_{13}(1720)$ is a
four-star resonance listed in PDG \cite{PDG}, and the later three resonances
can be tentatively identified with the two-star $S_{11}(1895)$, one-star
$P_{11}(2100)$, and one-star $P_{13}(2040)$ as listed in PDG \cite{PDG}. All
three above-threshold resonances quoted above are essential for achieving the
fit quality obtained in this work. Leaving out any one of them deteriorates the
fit quality considerably. The high-precision CLAS photoproduction data
\cite{CLAS09} constrain the masses of the above-threshold resonances very well.

We emphasize that, given the absence of spin-observable data, only the 
\emph{combined} analysis in the present work of recently obtained high-precision 
cross-section data across different reactions enabled us to impose sufficient 
constraints to unambiguously determine a minimum set of spin-1/2 and -3/2 
resonances. Incorporation of more resonances, especially of higher-spin 
resonances, requires experimental data on spin observables. 
Currently, the beam asymmetry and the beam-target asymmetry in 
$\gamma p \to \eta^\prime p$ are being measured by the CLAS Collaboration 
\cite{Collins09} and CBELSA/TAPS Collaboration \cite{Afzal12}, respectively.

In the free $\gamma p \to \eta' p$ reaction, there is a significant discrepancy
between the most recent CLAS \cite{CLAS09} and CBELSA/TAPS
\cite{CBELSA09} data. Since currently there is no clear reason to discard one
set in favor of the other, these two data sets lead to differences in the
extracted resonance parameters depending on which set is used for the analysis.
The major difference is in the extracted coupling strength of the $P_{11}(2130)$
resonance, where the CBELSA/TAPS data \cite{CBELSA09} yield a much larger coupling
than the CLAS data \cite{CLAS09} due to the larger cross sections
exhibited by the CBELSA/TAPS data at higher energies.

The quasi-free $\gamma p \to \eta' p$ and $\gamma n \to \eta' n$ reactions have
been also considered as a part of the combined analysis of the photon- and
nucleon-induced reactions. The latter (quasi-free) reaction helps to constrain
the set of above-threshold resonances.  Overall, these reactions were
reasonably well described by folding the cross sections of the corresponding
free processes with the Fermi distribution of the nucleon inside the deuteron.
The ratio of the neutron to proton electromagnetic couplings for the considered
resonances were extracted. Overall, the CBELSA/TAPS proton quasi-free data
\cite{CBELSA11} coincide with the proton free data within their uncertainties
for most of the angles and energies, although the quasi-free data are more in
line with the free CLAS data \cite{CLAS09} than with the free CBELSA/TAPS data
\cite{CBELSA09} at higher energies, where the effect of the Fermi motion of the
nucleon inside the deuteron is expected to have faded out. This might be an
indication of a possible problem with the CBELSA/TAPS proton free
data \cite{CBELSA09}. At lower energies, down to $W = 1.935$ GeV, the lowest
energy for which the proton free data exist, the data of Ref.~\cite{CBELSA11}
show no sign of the Fermi-motion effect, while the present model calculation
exhibits this effect up to $W \sim 2$ GeV.  Further studies are necessary to
identify the cause of this seeming discrepancy between the data and the model
result.

We found that the existing photon-induced reactions themselves can be described
by a set of three above-threshold resonances. However, the photoproduction data
alone cannot constrain the set of resonances uniquely. In the present analysis,
we also incorporate the $pp$ invariant mass
distribution data in the $NN \to NN\eta'$ reaction to obtain a restriction to a
single set of above-threshold resonances. The data clearly require the
above-threshold $P_{13}(2050)$ resonance. Furthermore, the $\eta'$ angular
distribution data in this reaction cannot be described adequately without the
sub-threshold $P_{13}(1720)$ resonance. Clearly, these findings illustrates
that meson productions in $NN$ collisions can help impose constraints on the
resonances. As pointed out in Ref.~\cite{NH06}, spin observables in $\eta'$
photoproduction, in particular the beam asymmetry, are much more sensitive to
the details of the model than are the cross sections. We also expect that the
analyzing power in $NN \to NN \eta'$ is sensitive to the excitation mechanism
of a given resonance, as is the case for the $NN \to NN\eta$ reaction
\cite{NSL02}. Also, in view of the contrasting results in $pp \to pp\eta$  and
$pp \to pp\eta'$ reactions shown in Fig.~\ref{fig:14}, it will be very
interesting to measure the $pp$ invariant mass distribution at an excess energy
of $Q \sim 45$ MeV in the latter reaction to learn more about the possible
production mechanism(s) in this reaction.

As has been pointed out in Refs.~\cite{NH04,NH06}, the determination of the 
coupling strength of $\eta'$ to nucleon is of special interest, particularly, in connection to the so-called nucleon spin crisis. In Ref.~\cite{NH06}, based on the then available data, we estimated
the upper limit of the $NN\eta'$ coupling constant to be
$g_{NN\eta'} \lesssim 2$. From the present analysis, with much
higher precision data, we now expect this coupling constant to be not much
larger than $g_{NN\eta'} \approx 1$, as can be seen from Table~\ref{tab:1}.

Finally, even though the hadronic final-state interactions in the reactions 
$\gamma N \to \eta^\prime N$ and $\pi N \to \eta^\prime N$ was not considered 
explicitly in the present work, it has been accounted for effectively through the 
generalized contact current of the photoprocess. In principle, the $\eta^\prime N$ 
FSI should be determined in a dynamical coupled channels approach, such as 
that of Ref.~\cite{Juelich12}. However, this is currently not an easy task in practice 
because the scarcity of the relevant data in the $\eta^\prime N$ channel restricts
severely the determination of the FSI in this channel with reasonable accuracy. 
In this connection, spin observables -- such as the target asymmetry in 
photoproduction -- may be of particular relevance to help determine the FSI 
\cite{NOH11}.

\begin{acknowledgments}
The authors thank V. Crede, I. Jaegle, B. Krusche, L. Tiator, and M. Williams
for fruitful discussions. F.H. is grateful to Profs. Zong-Ye Zhang and
Bing-Song Zou for their hospitality during his visit in the Institute of High
Energy Physics, Chinese Academy of Sciences, where part of this work was
completed. This work was supported by the FFE-COSY Grant No. 41788390
(COSY-58).
\end{acknowledgments}


\appendix

\section{Formalism}\label{sec:appA}

The formalism used in the present work is the same as that of
Refs.~\cite{NH04,NH06}. For completeness, we provide here a brief description
of this approach whose dynamical content is summarized by the graphs of
Figs.~\ref{fig:etaphotoprod} and \ref{fig:ppetaprod}.

\subsection{Free photoproduction}

\begin{figure}[tb]
\includegraphics[width=\columnwidth,angle=0,clip]{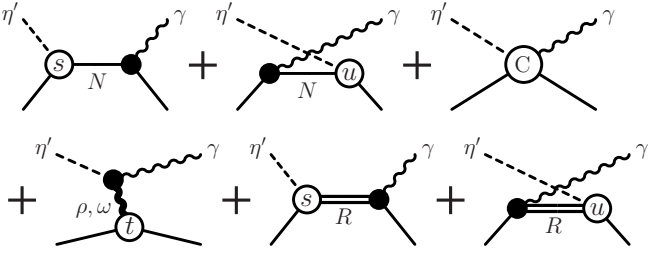}
\caption{\label{fig:etaphotoprod}
Diagrams contributing to $\gamma p \to \eta' p$. Time proceeds from right to left. The
intermediate baryon states are denoted $N$ for the nucleon, and $R$ for
the $S_{11}$ and $P_{11}$ resonances. The intermediate mesons in the $t$-channel are
$\rho$ and $\omega$. The external legs are labeled by the four-momenta of the respective
particles and the labels $s$, $u$, and $t$ of the hadronic vertices
correspond to the off-shell Mandelstam variables of the respective intermediate
particles. The three diagrams in the lower part of the diagram are transverse
individually; the three diagrams in the upper part are made gauge-invariant by an
appropriate choice (see text) of the contact current depicted in the top-right diagram.
The nucleonic current (nuc) referred to in the text corresponds to the top line of
diagrams; the meson-exchange current (mec) and resonance current contributions
correspond, respectively, to the leftmost diagram and the two diagrams on the right of
the bottom line of diagrams.}
\end{figure}

For the $\eta'$ photoproduction, we employ the tree graphs of
Fig.~\ref{fig:etaphotoprod} with form factors at the vertices to account for
their hadronic structure. The gauge invariance of this production current is
ensured by a phenomenological contact current that accounts for the effects of
the final-state interaction current not taken into account
explicitly \cite{HHK11,Huang12}. Following Refs.~\cite{HHK11,Huang12}, the generalized contact
current $M^\mu_c$ is chosen to be
\begin{equation}
M^\mu_c = \Gamma_{NN\eta^\prime}(q)C^\mu,
\label{eq:7}
\end{equation}
where $\Gamma_{NN\eta^\prime}$ stands for the $NN\eta^\prime$ vertex operator,
without the corresponding isospin operator and the form factor. The latter
enter the auxiliary current $C^\mu$ that is given by
\begin{equation}
C^\mu =  -e_f \frac{(2p'-k)^\mu}{u-p'^2}(f_u-\hat{F})
                 -e_i \frac{(2p+k)^\mu}{s-p^2}   (f_s-\hat{F}),
 \label{eq:8}
\end{equation}
with $k$, $p$, and $p^\prime$ denoting the four-momenta of the incoming
photon, initial nucleon and final nucleon, respectively, and
\begin{equation}
  \hat{F}= 1-\hat{h}\,\big(1-\delta_s f_s\big) \big(1-\delta_u f_u\big),
  \label{eq:9}
\end{equation}
with indices $x=s,u$ corresponding to the Mandelstam variables appropriate for
the respective kinematic situations depicted in Fig.~\ref{fig:etaphotoprod}.
The factors $\delta_x$ are unity if the corresponding channel contributes to
the reaction in question, and they are zero otherwise; $f_x$ denotes the
appropriate form factor. The parameter $\hat{h}$ may be an arbitrary (complex)
function, $\hat{h}=\hat{h}(s,u,t)$ which, in general, is subject to
crossing-symmetry constraints. (However, in the application discussed in this
work, we simply take $\hat{h}$ as a fit constant.)

The interaction Lagrangians as well as the form factors that provide the
meson-nucleon-baryon and photon-nucleon-baryon vertices involved in the
amplitudes $M^\mu_s$, $M^\mu_u$, $M^\mu_t$, and $M^\mu_c$ are given in
Appendix \ref{sec:appB}.

\subsection{Quasi-free photoproduction}\label{app:QF}

Following Refs.~\cite{ABHHM73,Krusche95,CBELSA11}, the quasi-free $\eta'$
photoproduction processes are described within a spectator model by folding the
cross sections for the corresponding free processes with the momentum
distribution of the nucleon inside the deuteron.

In the laboratory frame, the deuteron is at rest and the spectator nucleon
inside the deuteron is on its mass-shell and has three-momentum $\textbf{p}_s =
-\textbf{p}_N$. Therefore, the energy $E_N$ of the participant nucleon inside
the deuteron is given by
\begin{equation}\label{eq:a10}
E_N = M_d - \sqrt{m_s^2 + \textbf{p}_N^2},
\end{equation}
where $m_s$ denotes the mass of the spectator nucleon.

The invariant mass square of the $\gamma(k) + N(p_N) \to \eta^\prime(q) +
N(p'_N)$ subsystem, $Q^2 \equiv (k+p_N)^2$, can be expressed in the laboratory
frame, where the participant nucleon inside the deuteron has three-momentum
$\textbf{p}_N$, as
\begin{equation}\label{eq:a9}
Q^2(\textbf{p}_N) = E_N^2 - \textbf{p}_N^2 + 2E_\gamma\left(E_N - \textbf{p}_N \cdot \hat{\textbf{k}} \right),
\end{equation}
where $E_\gamma = (s - M_d^2)/(2\sqrt{s})$ denotes the incident photon energy
with $s$ denoting the invariant mass squared of the $\gamma d$ system and $M_d$ denoting the deuteron mass. The unit vector is given by $\hat{\textbf{k}} \equiv \textbf{k} / |\textbf{k}|$, where
$\textbf{k}$ denotes the three-momentum of the incident photon.

Of course, one must have $Q(\textbf{p}_N) \ge m_N + m_{\eta^\prime}$ for the
quasi-free $\gamma N \to \eta^\prime N$ process to take place. Together with
this condition, the invariant mass squared, $Q^2$, in Eq.~(\ref{eq:a9}) can be
expressed in terms of the four-momentum transfer square $t \equiv (p_d - p_s)^2
= p_N^2=E_N^2 - \textbf{p}_N^2$, where $p_d$ and $p_s$ stand for the
four-momenta of the deuteron and the spectator nucleon, respectively, as
\begin{align}\label{eq:a9b}
Q^2(\textbf{p}_N) & = t + 2E_\gamma\left(\sqrt{\textbf{p}_N^2 + t}  - \textbf{p}_N \cdot \hat{\textbf{k}} \right) \nonumber \\
& \equiv Q^2(t, \textbf{p}_N) \ .
\end{align}

The differential cross section of the quasi-free photoproduction process is,
then, approximated as
\begin{widetext}
\begin{equation}
\left. \frac{{\rm d}\sigma}{{\rm d}\Omega}\right |_{\rm quasi} \! \big(W, \theta\big) \,=\, \int {\rm d}^3 p_N \left| \Psi(\textbf{p}_N) \right|^2 \Theta\!\left(Q(t, \textbf{p}_N)-m_N - m_{\eta'}\right) \left. \frac{{\rm d}\sigma}{{\rm d}\Omega}\right |_{\rm free} \! \big(W'=Q(t, \textbf{p}_N), \theta\big),
\label{eq:QFXb}
\end{equation}
where $W\equiv Q(t=m_N^2, \textbf{p}_N=0)$.
 $\Theta(x)$ is the usual step function that is equal to 1 for $x\ge 0$ and 0
otherwise. It has been introduce for the sole purpose of making explicit that
$Q(t, \textbf{p}_N) \ge m_\eta' + m_N$ for the free $\gamma N \to \eta^\prime
N$ process to take place. $\theta$ is the scattering angle between outgoing
meson and incoming photon. $\Psi(\textbf{p}_N)$ is the deuteron wave function
in momentum space. ${\rm d}\sigma/{\rm d}\Omega |_{\rm free}$ is the
differential cross section for a free photoproduction process.

A variant of Eq.~(\ref{eq:QFXa}) is to restrict the participant nucleon to be
on its mass-shell in the argument of ${\rm d}\sigma/{\rm d}\Omega |_{\rm free}$
in Eq.~(\ref{eq:QFXb}) \cite{Krusche95,Anisovich09,CBELSA11}, i.e.,
\begin{equation}
\left. \frac{{\rm d}\sigma}{{\rm d}\Omega}\right |_{\rm quasi} \! \big(W, \theta\big) \,=\, \int {\rm d}^3 p_N \left| \Psi(\textbf{p}_N) \right|^2 \Theta\!\left(Q(t, \textbf{p}_N)-m_N - m_{\eta'}\right) \left. \frac{{\rm d}\sigma}{{\rm d}\Omega}\right |_{\rm free} \! \big(W'=Q(m_N^2, \textbf{p}_N), \theta\big) \ .
\label{eq:QFXa}
\end{equation}
\end{widetext}
There are at least two reasons for this on-shell restriction. One is the fact
that the free cross sections that enter the equations above are ``on shell". The other is that by
restricting $t=m_N^2$, the effect of the Fermi folding is to smear out the free
cross section keeping the centroid position, corresponding to
$d\sigma/d\Omega{\big |}_{\text{free}}(W, \theta)$, not to be shifted. The
latter feature seems to better reproduce the quasi-free $\eta$ photoproduction
data \cite{Krusche95,Anisovich09}.

In the present work we employ Eq.~(\ref{eq:QFXa}), but also obtained results
using Eq.~(\ref{eq:QFXb}) and we find little difference between the two
prescriptions.

\subsection{Hadronic production}

\begin{figure}[tb]
\centering\includegraphics[width=0.8\columnwidth,angle=0,clip=]{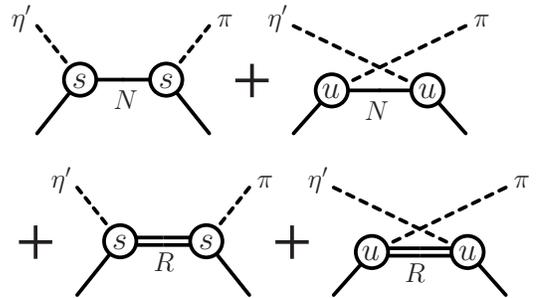}
\caption{\label{fig:diagram_pieta}
Feynman diagrams contributing to $\pi N \to  \eta' N$. The notation is the same
as in Fig.~\ref{fig:etaphotoprod}. }
\end{figure}

The $\pi N \to \eta' N$ reaction is described here within a tree-level
approximation, analogous to the description of $\pi N \to \eta N$ in
Ref.~\cite{NOH11}.  We take into account the nucleonic and resonance
contributions as depicted in Fig.~\ref{fig:diagram_pieta}. In principle, one
could also include the $t$-channel diagrams such as the (rank-two) tensor meson
$a_2(1320)$ exchange, whose decay branching ratio to $\eta'\pi$ is quoted to be
$BR(a_2\to \eta'\pi) = 5.3 \pm 0.9 \times 10^{-3}$ \cite{PDG}. We have not
considered such contributions in the present work since we do not expect that
including them would alter our results in any significant manner.
The propagators, vertices, and form factors necessary for calculating the
Feynman diagrams in Fig.~\ref{fig:diagram_pieta} are given in Appendix \ref{sec:appB}.

\begin{figure}[tb]
\includegraphics[width=\columnwidth,angle=0,clip]{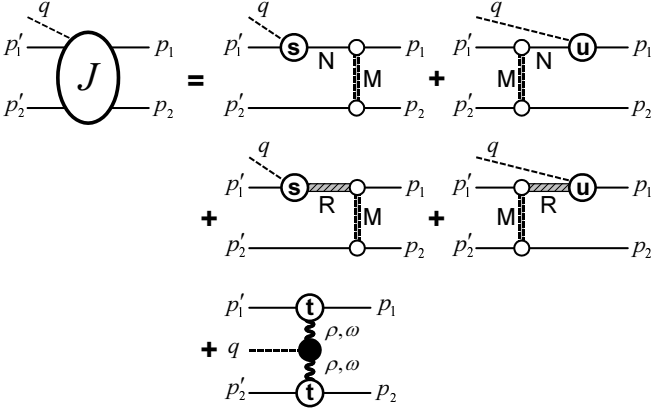}
\caption{\label{fig:ppetaprod}
Basic production mechanisms for $NN\to NN \eta'$. Time proceeds from right to left. The
full amplitude, with additional initial- and final-state contributions, is given by
Eq.~(\ref{eq:C1}). As in Fig.~\ref{fig:etaphotoprod}, $N$ and $R$
denote the intermediate nucleon and resonances, respectively, and $M$ incorporates
all exchanges of mesons $\pi$, $\eta$, $\rho$, $\omega$, $\sigma$, and $a_0$
($\equiv\delta$) for the nucleon graphs and $\pi$, $\rho$, and $\omega$ for the resonance
graphs. External legs are labeled by the four-momenta of the respective particles; the
hadronic vertices $s$, $u$, and $t$ here correspond to the same
kinematic situations, respectively, as those identified similarly in
Fig.~\ref{fig:etaphotoprod}. The nucleonic (nuc), resonance, and meson-exchange (mec)
contributions
referred to in the text correspond, respectively, to the first, second, and third lines of
the diagrams on the right-hand side.}
\end{figure}

The hadronic reaction $NN \to NN\eta'$ is described according to the model put forward in
Refs.~\cite{NADS99,NSL02,NOH11}. The DWBA amplitude $M$ for this process is given by
\begin{equation}
M = (1 + T_f G_f) J (1 +  G_i T_i),
  \label{eq:C1}
\end{equation}
where $T_{n}$, with $n=i,f$, denotes the $NN$ $T$-matrix interaction in the
initial ($i$) or final ($f$) state, and $G_n$ is the corresponding two-nucleon
propagator (which absorbs the factor $i$ found in the DWBA formula given in
Ref.~\cite{NSL02}).  $J$ sums up the basic $\eta'$ production mechanisms
depicted in Fig.~\ref{fig:ppetaprod}. The interaction Lagrangians as well as
the form factors necessary for constructing the basic production amplitude $J$
are given in Appendix B. In the absence of models capable of providing a
reliable off-shell $NN$ initial state interaction (ISI), we consider it only in
the on-shell approximation following Refs.~\cite{HN99,NSL02}. This was shown to
be a reasonable approximation for calculating cross sections \cite{HN99}.
For the on-shell $NN$ interaction, we consider the phase-shifts and
inelasticities from the SAID partial-wave analysis \cite{SAID}. All the partial
waves up to total angular momentum $J=7$ are considered.
The $NN$ FSI is generated using the Paris potential \cite{Paris}
where the Coulomb interaction is taken into account fully as described in Ref.~\cite{NADS99}.
We also follow Ref.~\cite{HN92} to convert the Paris $NN$ interaction, which obeys
the non-relativistic Lippman-Schwinger equation, to the one obeying the three-dimensionally reduced version (a la Blankenbecler-Sugar) of the relativistic Bethe-Salpeter
equation in order to be consistent with the relativistic covariant approach used in the
present work. We also use the Blankenbecler-Sugar propagator for the two-nucleon
propagator $G_f$ in Eq.~(\ref{eq:C1})
for consistency.

\section{Lagrangians, Form Factors, and Propagators}\label{sec:appB}

The interaction Lagrangian used to construct our model for the basic production
amplitudes is given below. For further convenience, we define the operators
\begin{equation}
\Gamma^{(+)} = \gamma_5
\qquad\text{and}\qquad
\Gamma^{(-)} = 1.
\end{equation}

\subsection{Electromagnetic interaction Lagrangians}

\begin{align}
{\cal L}_{NN\gamma} =& \, -e \bar N \left[ \left(\hat{e} \gamma^\mu
- \frac{\hat{\kappa}}{2M_N}\sigma^{\mu\nu}\partial_\nu \right) A_\mu
\right] N, \label{NNr} \\[2pt]
{\cal L}_{\eta'\rho\gamma} =&\; e\frac{g_{\eta'\rho\gamma}}{m_{\eta'}}
\varepsilon_{\alpha\mu\lambda\nu} \left(\partial^\alpha
A^\mu\right) \left(\partial^\lambda \eta' \right) \rho_3^\nu, \label{L_getarho} \\[2pt]
{\cal L}_{\eta'\omega\gamma} =&\; e\frac{g_{\eta'\omega\gamma}}{m_{\eta'}}
\varepsilon_{\alpha\mu\lambda\nu} \left(\partial^\alpha
A^\mu\right) \left(\partial^\lambda \eta'\right) \omega^\nu, \label{L_getaome}
\end{align}
where $e$ stands for the elementary charge unit, and $\hat e \equiv (1 +
\tau_3)/2$ and $\hat\kappa \equiv \kappa_p(1 + \tau_3)/2 + \kappa_n(1 -
\tau_3)/2$, with the anomalous magnetic moments $\kappa_p=1.793$ of the proton
and $\kappa_n=-1.913$ of the neutron; $M_N$ stands for the nucleon mass.
$F_{\mu\nu}\equiv \partial_\mu A_\nu - \partial_\nu A_\mu$ with $A_\mu$
denoting the electromagnetic field and $\varepsilon_{\alpha\mu\lambda\nu}$ is
the totally antisymmetric Levi-Civita tensor with $\varepsilon^{0123}=+1$. The
meson-meson electromagnetic transition coupling constants in the above
Lagrangians, $g_{\eta'\rho\gamma}=1.25$ and $g_{\eta'\omega\gamma}=0.44$, are
extracted from a systematic analysis of the radiative decay of pseudoscalar and
vector mesons based on flavor $SU(3)$ symmetry considerations in conjunction
with vector-meson dominance arguments \cite{NADS99}; their signs are inferred
in conjunction with the sign of the coupling constant $g_{\pi v\gamma}$
($v=\rho,\omega$) determined from a study of pion photoproduction in the 1 GeV
energy region \cite{GdG93}. The resulting $\eta^\prime v\gamma$ vertex is
multiplied by the form factor $\tilde{f}_v(t)$ which describes the off-shell
behavior of the intermediate vector meson with squared momentum transfer
$t=(p-p')^2$ (cf. fourth diagram in Fig.~\ref{fig:etaphotoprod}). We use the
dipole form
\begin{equation}
  \tilde{f}_v(t) = \left( \frac{{\Lambda^*_v}^2}{{\Lambda^*_v}^2-t} \right)^2.
  \label{eq:dipoleFF}
\end{equation}
The cutoff $\Lambda^*_v$, taken to be identical for both $\rho$ and $\omega$, is a fit parameter.

The resonance-nucleon-photon-transition Lagrangians are
\begin{subequations}
\begin{align}
\mathcal{L}^{(\frac{1}{2}^\pm)}_{RN\gamma} = & \;  e \frac{g_{RN\gamma}^{(1)}}{2M_N} \bar R \Gamma^{(\mp)}
\sigma_{\mu\nu}\left(\partial^\nu A^\mu\right) N + {\rm H.c.}, \label{1hfNr} \\[1pt]
\mathcal{L}^{(\frac{3}{2}^\pm)}_{RN\gamma} = & \, -ie\frac{g^{(1)}_{RN\gamma}}{2M_N} {\bar{R}^\mu} \gamma_\nu
\Gamma^{(\pm)} F^{\mu\nu} N \nonumber \\
&\, + e\frac{g^{(2)}_{RN\gamma}}{4M^2_N} \bar{R}^\mu \Gamma^{(\pm)} F^{\mu\nu}
\partial_\nu N + {\rm H.c.},
\label{3hfNr}
\end{align}
\end{subequations}
where the superscript of $\mathcal{L}_{RN\gamma}$ denotes the spin and parity
of the resonance $R$. The  coupling constants $g^{(i)}_{RN\gamma}$ ($i=1,2$) are
fit parameters.

\subsection{Hadronic interaction Lagrangians}\label{app:hadronLagrange}

In this subsection, we use $S$ $(=\sigma, \vec a_0)$, $P$ $(=\eta, \vec\pi)$,
and $V_\mu$ $(=\omega_\mu, \vec\rho_\mu)$ to denote the scalar, pseudoscalar,
and vector meson fields, respectively. The vector notation refers to the
isospin space. For isovector mesons, $S \equiv \vec S \cdot \vec \tau$, $P
\equiv \vec P \cdot \vec \tau$, and $V_\mu \equiv \vec V_\mu \cdot \vec \tau$.

The Lagrangians for meson-nucleon interactions are
\begin{subequations}
\begin{align}
{\cal L}_{NNS} &=  g_{NNS}^{} \,\bar N N S ,
\label{NNS}
\displaybreak[0]\\
{\cal L}_{NNP} &=
- g_{NNP}^{}\, \bar N \left\{ \Gamma^{(+)} \left[ i\lambda + \frac{1 - \lambda}{2M_N}\, \fs{\partial} \right] P \right\} N ,
\label{NNP} \displaybreak[0]\\
{\cal L}_{NNV} &=  - g_{NNV}^{} \,
\bar N \left\{ \left[ \gamma^\mu - \kappa_V^{} \frac{\sigma^{\mu\nu}\partial_\nu}{2M_N} \right] V_\mu \right\} N ,
\label{NNV}
\end{align}
\end{subequations}
where the parameter $\lambda$ was introduced in $\mathcal{L}_{NNP}$ to
interpolate between the pseudovector ($\lambda=0$) and the pseudoscalar
($\lambda=1$) couplings.
  The $NN\eta'$ coupling constant,
$g_{NN\eta'}$, is a fit parameter. All the other coupling constants in the
above Lagrangians are taken from Ref.~\cite{Bonn}, with the exception of
$g_{NN\omega}=10$ as explained in Ref.~\cite{NSL02}.

The $\eta' vv$ ($v=\rho,\omega$) Lagrangians are
\begin{align}
{\cal L}_{\eta'\rho\rho} =&\; \frac{g_{\eta'\rho\rho}}{2m_{\eta'}}
\varepsilon_{\alpha\mu\lambda\nu} \left(\partial^\alpha
\vec{\rho}^{\,\mu}\right)\left(\partial^\lambda \eta' \right)\cdot \vec{\rho}^{\,\nu}, \label{L_getarho2} \\[2pt]
{\cal L}_{\eta'\omega\omega} =&\; \frac{g_{\eta'\omega\omega}}{2m_{\eta'}}
\varepsilon_{\alpha\mu\lambda\nu}  \left(\partial^\alpha
\omega^\mu\right)\left(\partial^\lambda \eta'\right)  \omega^\nu. \label{L_getaome2}
\end{align}
The coupling constants, $g_{\eta'\rho\rho}^{}=5.51$ and
$g_{\eta'\omega\omega}^{}=5.42$, are obtained from a systematic analysis of the
radiative decay of pseudoscalar and vector mesons based on $SU(3)$ symmetry
considerations;  their signs are inferred, in conjunction with vector-meson
dominance assumptions, from the sign of the coupling constant $g_{\pi v\gamma}$
($v=\rho,\omega$) determined from a study of pion photoproduction in the 1 GeV
energy region \cite{GdG93}.

\begin{widetext}

For nucleon resonances,
\begin{subequations}
\begin{align}
{\cal L}^{(\frac{1}{2}^\pm)}_{RNP} &=
- g_{RNP}^{} \bar R  \left\{ \Gamma^{(\pm)} \left[ \pm i\lambda +
  \frac{1 - \lambda}{M_R\pm M_N}\, \fs{\partial} \right]P\right\} N + \hc,
\label{PNR}
\displaybreak[0]\\
{\cal L}^{(\frac{1}{2}^\pm)}_{RNV} &=
\frac{g_{RNV}}{2M_N} \bar R \Gamma^{(\mp)} \sigma_{\mu\nu}\left(\partial^\nu V^\mu\right) N + \hc,
\label{VNR}
\displaybreak[0]\\
\mathcal{L}^{(\frac{3}{2}^\pm)}_{RNP} &=
\frac{g_{RNP}^{}}{M_P} {\bar{R}_\mu} \Gamma^{(\mp)} (\partial^\mu P) N + \hc,
 \label{PNR32}
\displaybreak[0]\\
\mathcal{L}^{(\frac{3}{2}^\pm)}_{RNV} &=
- i\frac{g^{(1)}_{RNV}}{2M_N} {\bar{R}_\mu}  \gamma_\nu \Gamma^{(\pm)} V^{\mu\nu} N
+ \frac{g^{(2)}_{RNV}}{4M^2_N} \bar{R}_\mu  \Gamma^{(\pm)} V^{\mu\nu} \partial_\nu N
\mp \frac{g^{(3)}_{RNV}}{4M^2_N} \bar{R}^\_\mu \Gamma^{(\pm)} \left(\partial_\nu V^{\mu\nu}\right) N + \hc.
\label{VNR32}
\end{align}
\end{subequations}
where the Lagrangian (\ref{PNR}) contains the pseudoscalar-pseudovector mixing
parameter $\lambda$, similar to Eq.~(\ref{NNP}).

\end{widetext}

Each hadronic vertex obtained from the interaction Lagrangians given in this
subsection is multiplied by a phenomenological cutoff function
\begin{equation}
f(p'^2, p^2, q^2) = f_B(p'^2) f_B(p^2) f_M(q^2)  ,
\label{ffhadron}
\end{equation}
where $p'$ and $p$ denote the four-momenta of the two baryons, and $q$ is the
four-momentum of the meson at the three-point vertex. Here, we use
\begin{equation}
f_B(x) =\frac{\Lambda_B^4}{\Lambda_B^4+\left( x-M_B^2 \right)^2}  ,
\label{ffbaryon}
\end{equation}
where the cutoff $\Lambda_B=1200$ MeV is taken the same for all the baryons
$B$, and $f_M(q^2)$ is given by
\begin{equation}
f_M(q^2) = \left( \frac{\Lambda_M^2-m_M^2}{\Lambda_M^2-q^2} \right)^n  ,
\label{ffmeson}
\end{equation}
with $n=1$ for a scalar or a pseudoscalar meson and $n=2$ for a vector meson.
$m_M$ denotes the mass of meson $M$. The values of $\Lambda_M$ are taken the
same as those used in Ref.~\cite{NSL02}.

\subsection{Energy-dependent resonance widths}\label{sec:widths}

Our formalism is adapted to accommodate energy-dependent resonance widths with
the appropriate threshold behavior.

For a spin-1/2 resonance propagator, we use the ansatz
\begin{equation}
S_{1/2}(p) = \frac{1}{\fs{p}-m_R
+\frac{i}{2}\Gamma}=\frac{\fs{p}+m_R}{p^2-m_R^2+\frac{i}{2}(\fs{p}+m_R)\Gamma},
\label{eq:spin1}
\end{equation}
where $m_R$ is the mass of the resonance with four-momentum $p$. $\Gamma$ is
the width function whose functional behavior will be given below.

For spin-3/2, the resonant propagator reads in a schematic matrix notation
\begin{equation}
S_{3/2}(p)=\left[(\fs{p}-m_R)g-i\frac{\Delta}{2}\Gamma\right]^{-1}\Delta.
\label{eq:spin3}
\end{equation}
All indices are suppressed here, i.e., $g$ is the metric tensor and $\Delta$ is
the Rarita--Schwinger tensor written in full detail as
\begin{equation}
\Delta^{\mu\nu}_{\beta\alpha}=
-g^{\mu\nu}\delta_{\beta\alpha}+\frac{1}{3}\gamma^\mu_{\beta\varepsilon}\gamma^\nu_{\varepsilon\alpha}
        + \frac{2p^\mu p^\nu}{3m_R^2}\delta_{\beta\alpha}
                 +\frac{\gamma^\mu_{\beta\alpha} p^\nu-p^\mu\gamma^\nu_{\beta\alpha}}{3m_R},
\label{eq:RStensor}
\end{equation}
where $\beta$, $\alpha$, and $\varepsilon$ enumerate the four indices of the
$\gamma$-matrix components (summation over $\varepsilon$ is implied). The
inversion in (\ref{eq:spin3}) is to be understood on the full 16-dimensional
space of the four Lorentz indices and the four components of the gamma
matrices.

In both cases, we write the width $\Gamma$ as a function of $W=\sqrt{s}$
according to
\begin{equation}
\Gamma(W) = \Gamma_R \sum_{i=1} \beta_i \hat{\Gamma}_i(W),
\label{eq:17}
\end{equation}
where the sums over $i$ accounts for decays of the
resonance into two- or three-hadron channels and into radiative
decay channels. The total static resonance width is denoted by $\Gamma_R$ and
the numerical factors $\beta_i$ describes the branching ratios into the various
decay channels, i.e.,
\begin{equation}
\sum_{i=1}^N \beta_i  =1.
\end{equation}
Similar to Refs.~\cite{walker,arndt90,lvov97,drechsel}, we parameterize the
width functions $\hat{\Gamma}_i$ (which is normalized to unity at $W=m_R$)
to provide the correct respective threshold behaviors. The details may be found in
Ref.~\cite{NH06}.



\begin{thebibliography}{99}


\bibitem{NH04}K. Nakayama and H. Haberzettl, Phys. Rev. C \textbf{69}, 065212 (2004).

\bibitem{Capstick1}S. Capstick and N. Isgur, Phys. Rev. D \textbf{34}, 2809 (1986);
S. Capstick and W. Roberts, \textit{ibid.} \textbf{47}, 1994 (1993); \textbf{49}, 4570 (1994); \textbf{57}, 4301 (1998); \textbf{58}, 074011 (1998).

\bibitem{CLAS06}M. Dugger \textit{et al.}, Phys. Rev. Lett. \textbf{96}, 062001 (2006); Erratum-ibid. \textbf{96}, 169905 (2006).

\bibitem{CLAS09}M. Williams \textit{et al.}, Phys. Rev. C \textbf{80}, 045213 (2009).

\bibitem{CBELSA09}V. Crede \textit{et al.}, Phys. Rev. C \textbf{80}, 055202 (2009).

\bibitem{CBELSA11}I. Jaegle \textit{et al.}, Eur. Phys. J. A \textbf{47}, 11 (2011).

\bibitem{COSY11-10a}P. Klaja \textit{et al.}, Phys. Lett. B \textbf{684}, 11 (2010).

\bibitem{ppetapdata}F. Hibou \textit{et al.}, Phys. Lett. B \textbf{438}, 41 (1998); P. Moskal \textit{et al.}, Phys. Rev. Lett. \textbf{80}, 3202 (1998); P. Moskal \textit{et al.}, Phys. Lett. B \textbf{474}, 416 (2000).

\bibitem{Balestra}F. Balestra \textit{et al.}, Phys. Lett. B \textbf{491}, 29 (2000).

\bibitem{Khoukaz}A. Khoukaz \textit{et al.}, COSY-11 Collaboration, Eur. Phys. J. A \textbf{20}, 345 (2004).

\bibitem{COSY11-10b}J. Klaja \textit{et al.}, Phys. Rev. C \textbf{81}, 035209 (2010).

\bibitem{NH06}K. Nakayama and H. Haberzettl, Phys. Rev. C \textbf{73}, 045211 (2006).

\bibitem{Collins09}
P.~Collins, B.~G.~Ritchie, M.~Dugger, E.~Pasyuk, F.~Klein
\textit{et al.}, to be published.

\bibitem{Afzal12}
F. N. Afzal (CBELSA/TAPS Collaboration), 
EPJ Web of Conferences \textbf{37}, 09001 (2012).

\bibitem{ZZ11}Xian-Hui Zhong and Qiang Zhao, Phys. Rev. C \textbf{84}, 065204 (2011).

\bibitem{PDG}J. Beringer \textit{et al.} (Particle Data Group), Phys.\ Rev.\ D \textbf{86}, 010001 (2012).

\bibitem{SHKM10}A. Sibirtsev, J. Haidenbauer, S. Krewald, and U.-G. Mei\ss ner, Eur. Phys. J. A \textbf{46}, 359 (2010).

\bibitem{NOH11}K. Nakayama, Yongseok Oh, H. Haberzettl, J. Korean Phys. Soc. \textbf{59}, 224 (2011).

\bibitem{CYTVD03}W. T. Chiang, S. N. Yang, L. Tiator, M. Vanderhaegen, and D. Drechsel, Phys. Rev. C \textbf{68}, 045202 (2003).

\bibitem{Krusche95}B. Krusche \textit{et al.}, Phys. Lett. B \textbf{358}, 40 (1995).

\bibitem{Anisovich09}A. V. Anisovich \textit{et al.}, Eur. Phys. J. A \textbf{41}, 31 (2009).

\bibitem{COSYTOF03}M. Abdel-Bary \textit{et al.}, (COSY-TOF Collaboration), Eur. Phys. J. A \textbf{16}, 127 (2003).

\bibitem{COSY1104}P. Moskal \textit{et al.}, Phys. Rev. C \textbf{79}, 015208 (2009); P. Moskal \textit{et al.}, Phys. Rev. C \textbf{69}, 025203 (2004).

\bibitem{NHHS03}K. Nakayama, J. Haidenbauer, C. Hanhart, and J. Speth, Phys. Rev. C \textbf{68}, 045201 (2003).

\bibitem{Delof}A. Deloff, Phys. Rev. C \textbf{69}, 035206 (2004).

\bibitem{Dufey68}J. P. Dufey, B. Gobbi, M. A. Pouchon, A. M. Cnops, G. Finocchiaro, J. C. Lassalle, P. Mittner, and A. M\"uller, Phys. Lett. B \textbf{26}, 410 (1968); M. Basile {\it et al.}, Nuovo Cim. A \textbf{3}, 371 (1971); M. Basile {\it et al.}, Nucl. Phys. B \textbf{33}, 29 (1971).

\bibitem{Rader72}P. K. Rader {\it et al.}, Phys. Rev. D \textbf{6}, 3059 (1972); J. Bensinger, A. R. Erwin, M. A. Thompson, and W. D. Walker, Phys. Lett. B \textbf{33}, 505 (1970); R. J. Miller, S. Lichtman, and R. B. Willmann, Phys. Rev. \textbf{178}, 2061 (1969).

\bibitem{Baldini88}A. Baldini, V. Flaminio, W. G. Moorhead, and D. R. O. Morrison, Total Cross-Sections for Reactions of High Energy Particles, Landolt-B\"ornstein, edited by H. Schopper (Springer, Berlin, 1988), Vol I/12a.

\bibitem{Cao08}Xu Cao and Xi-Guo Lee, Phys. Rev. C \textbf{78}, 035207 (2008).

\bibitem{NSL02}K. Nakayama, J. Speth, and T.-S. H. Lee, Phys. Rev. C \textbf{65}, 045210 (2002).

\bibitem{Juelich12}D. R\"onchen, M. D\"oring, F. Huang, H. Haberzettl, J. Haidenbauer, C. Hanhart, S. Krewald, U.-G. Mei{\ss}ner, and  K. Nakayama,  Eur. Phys. J. A {\textbf 49}, 44 (2013).

\bibitem{HHK11}H. Haberzettl, F. Huang, and K. Nakayama, Phys. Rev. C \textbf{83}, 065502 (2011).

\bibitem{Huang12}F. Huang, M. D\"oring, H. Haberzettl, J. Haidenbauer, C. Hanhart, S. Krewald, U.-G. Mei{\ss}ner, and K. Nakayama, Phys. Rev. C \textbf{85}, 054003 (2012).

\bibitem{ABHHM73}P. Benz \textit{et al.}, Nucl. Phys. B \textbf{65}, 158 (1973).

\bibitem{NADS99}K. Nakayama, H. F. Arellano, J. W. Durso, and J. Speth, Phys. Rev. C \textbf{61}, 024001 (1999).

\bibitem{HN99}Ch. Hanhart and K. Nakayama, Phys. Lett. B \textbf{454}, 176 (1999).

\bibitem{SAID} SAID Data Analysis Center, Institute for Nuclear Studies, The George Washington University, Washington, DC 20052, USA. \url{http://gwdac.phys.gwu.edu}.

\bibitem{Paris}M. Lacombe, B. Loiseau, J. M. Richard, R. Vinh Mau, J. C{\^o}t\'e, P. Pir\`es, and R. de Tourreil, Phys. Rev. C \textbf{21}, 861 (1980).

\bibitem{HN92}V. Herrmann and K. Nakayama, Phys. Rev. C \textbf{46}, 2199 (1992).

\bibitem{GdG93}H. Garcilazo and E. Moya del Guerra, Nucl. Phys. A \textbf{562}, 521 (1993).

\bibitem{Bonn}R. Machleidt, Adv. Nucl. Phys. \textbf{19}, 189 (1989).

\bibitem{walker}R. L. Walker, Phys. Rev. \textbf{182}, 1729 (1969).

\bibitem{arndt90}R. A. Arndt, R. L. Workman, Z. Li, and L. D. Roper, Phys. Rev. C \textbf{42}, 1864 (1990).

\bibitem{lvov97}A. I. L'vov, V. A. Petrun'kin, and M. Schumacher, Phys. Rev. C \textbf{55}, 359 (1997).

\bibitem{drechsel}D. Drechsel, O. Hanstein, S. S. Kamalov, and L. Tiator, Nucl. Phys. A \textbf{645}, 145 (1999).


\end{thebibliography}
\end{document}